\documentclass[12pt]{article}

\usepackage{amsmath}
\usepackage{graphicx,psfrag,epsf}
\usepackage[ruled,vlined]{algorithm2e}

\usepackage{enumerate}
\usepackage{setspace}
\usepackage{amssymb}
\usepackage{bm}
\usepackage{enumitem}
\usepackage{booktabs}
\usepackage{multirow}
\usepackage{natbib}

\usepackage{xurl}
\usepackage[colorlinks,allcolors=blue]{hyperref} 
\usepackage{caption}
\usepackage{subcaption}
\newcommand{\blind}{0}

\graphicspath{ {./figures/} }

\begin{document}

\def\spacingset#1{\renewcommand{\baselinestretch}%
{#1}\small\normalsize} \spacingset{1}

\newcommand{\supplementarysection}{%
  \setcounter{figure}{0}
  \let\oldthefigure\thefigure
  \renewcommand{\thefigure}{S\oldthefigure}
  \section*{Supplement}
}


\if0\blind
{
  \title{\bf Ultra-efficient MCMC for Bayesian longitudinal functional data analysis}
  \author{Thomas Y. Sun\thanks{PhD student, Department of Statistics, Rice University (\url{tys1@rice.edu}).}, Daniel R. Kowal\thanks{Dobelman Family Assistant Professor, Department of Statistics, Rice University (\url{daniel.kowal@rice.edu}). This material is based upon work supported by the National Science Foundation (SES-2214726).}\hspace{.2cm}}
  \maketitle
} \fi

\if1\blind
{
  \bigskip
  \bigskip
  \bigskip
  \begin{center}
    {\LARGE\bf Title}
\end{center}
  \medskip
} \fi

\bigskip
\begin{abstract}
Functional mixed models are widely useful for regression analysis with dependent functional data, including longitudinal functional data with scalar predictors. However, existing algorithms for Bayesian inference with these models only provide either scalable computing or accurate approximations to the posterior distribution, but not both. We introduce a new MCMC sampling strategy for highly efficient and fully Bayesian regression with longitudinal functional data. Using a novel blocking structure paired with an orthogonalized basis reparametrization, our algorithm jointly samples the fixed effects regression functions together with all subject- and replicate-specific random effects functions. Crucially, the joint sampler optimizes sampling efficiency for these key parameters while preserving computational scalability. Perhaps surprisingly, our new MCMC sampling algorithm even surpasses state-of-the-art algorithms for frequentist estimation and variational Bayes approximations for functional mixed models---while also providing accurate posterior uncertainty quantification---and is orders of magnitude faster than existing Gibbs samplers. Simulation studies show improved point estimation and interval coverage in nearly all simulation settings over competing approaches. We apply our method to a large physical activity dataset to study how various demographic and health factors associate with intraday activity. 
\end{abstract}

\noindent%

\doublespacing
\section{Introduction}
\label{sec:intro}

Functional data analysis (FDA) refers to the statistical analysis of data objects observed over a continuum, such as time or space, typically at high resolutions. FDA has been applied in a variety of important  areas, including climate data \citep{besse_autoregressive_2000}, electricity prices \citep{liebl_modeling_2013}, COVID-19 dynamics \citep{boschi_functional_2021}, and many others. FDA is particularly challenging when the functional data are dependent, which requires more sophisticated statistical models and more intensive computations. We focus on regression analysis with \emph{longitudinal functional data}, which presents the simultaneous challenges of (i) within-curve dependencies, (ii) groupings among repeated (functional) measurements, and (iii) associations with (possibly many) scalar covariates. 

To meet these challenges, \emph{functional mixed models} (FMMs) have emerged as a powerful modeling tool. FMMs combine FDA and traditional mixed effects models to provide regression analysis of functional data in the presence of structured dependence. However, longitudinal functional datasets are often massive, with millions   or billions of high-resolution measurements \citep{doherty_large_2017, lee_bayesian_2019}. Thus, recent work has increasingly prioritized the \emph{computational scalability} of FMMs. Naturally, the attendant computational burdens are exacerbated by the complexity of the statistical model, including various covariance structures and (possibly many) scalar covariates. As a result, there is high demand for inferential algorithms for FMMs that both (i) scale to massive datasets and (ii) 
maintain these essential modeling capabilities. 

\begin{figure}[h]
\centering
\includegraphics[width=1\textwidth]{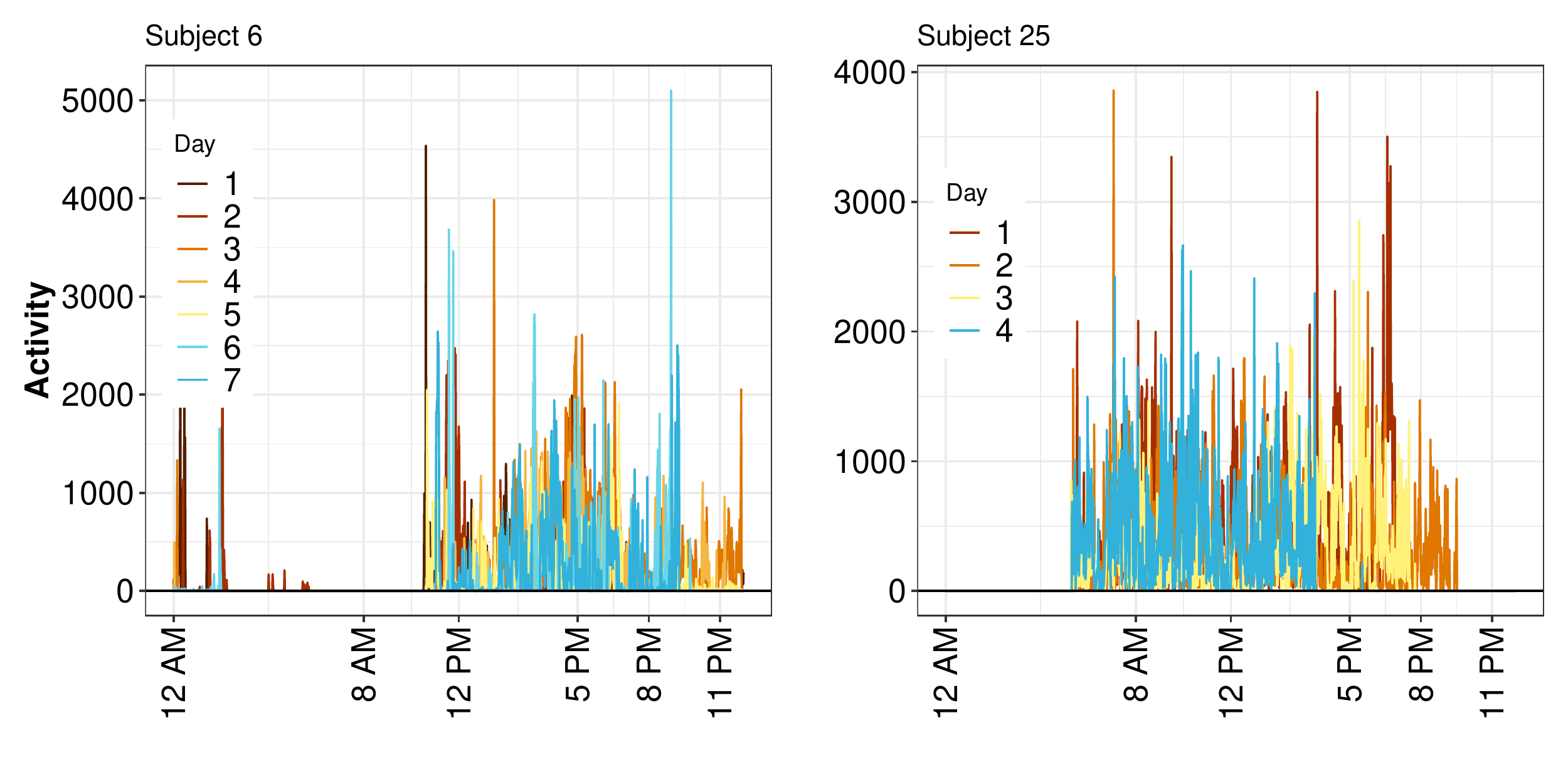}
\caption{Minute-by-minute physical activity (PA) measurements for two subjects in the NHANES study. The PA observations are noisy, autocorrelated, and exhibit considerable variation both between subjects and between days for a given subject.}
  \label{fig:actex}
\end{figure}

These challenges are exemplified in our motivating application, the 2005-2006 National Health and Nutrition Examination Survey (NHANES) physical activity (PA) dataset. High resolution (minute-by-minute) PA data (Figure~\ref{fig:actex}) are recorded across multiple days for each participant using hip-worn accelerometry devices and linked to a questionnaire that contains subject-level demographic and health information (Table~\ref{tab:variables}). 
Figure \ref{fig:actex} shows the PA levels of two randomly chosen subjects for every day they wore the device. Key aspects of the data become evident from this plot: the observations are high-resolution, noisy, autocorrelated, and exhibit considerable variation both between subjects and between days for a given subject. Notably, there are more than 72 million PA measurements on more than 10,000 participants in the study \citep{leroux_organizing_2019}. The goal is to link time-of-day PA with important health and demographic variables, while appropriately accounting for the prominent and complex dependencies among these high-dimensional data.


To enable statistical modeling and inference, we represent the PA as longitudinal functional data: the PA measurements are \emph{functions} of time-of-day and the days are \emph{repeated measurements} of these functional data for each subject.  More generally, let $Y_{i,j}(\tau)$ denote functional observations on a compact domain $\tau \in \mathcal{T}$ (i.e., time-of-day) for within-subject replications $j=1,\ldots,m_i$ (i.e., days) and subjects $i=1,\ldots,n$, with $M=\sum_{i=1}^n m_i$. The functional data are linked to $L$ scalar covariates $\bm x_{i,j} = (x_{i,j,1},\ldots, x_{i,j,L})'$ for subject $i$ and replicate $j$. We study the following FMM for longitudinal functional regression: 
\begin{equation}    
\begin{aligned}
Y_{i,j}(\tau) = \tilde{\alpha}_0(\tau) + \sum_{\ell=1}^{L} x_{i,j,\ell}\tilde{\alpha}_{\ell}(\tau) + \tilde{\gamma}_{i}(\tau) + \tilde{\omega}_{i,j}(\tau) + \epsilon_{i,j}\left(\tau\right), \quad \epsilon_{i,j}\left(\tau\right) \overset{iid}{\sim} N(0, \sigma^2_{\epsilon}). \label{fmm}
\end{aligned}
\end{equation}
Model \eqref{fmm} extends traditional mixed effects models for longitudinal data to the functional data setting by allowing the \emph{fixed effects} $\{\tilde \alpha_\ell(\cdot)\}$ and the \emph{random effects} $\{\tilde \gamma_i(\cdot), \tilde \omega_{i,j}(\cdot)\}$ to vary as functions of $\tau$. First, the fixed effects $\{\tilde \alpha_\ell(\cdot)\}$ are regression coefficient functions and describe the linear associations between $\bm x_{i,j}$ and $Y_{i,j}(\tau)$ at each $\tau$. Next, the  random effects $\{\tilde \gamma_i(\cdot)\}$ are shared among all functional observations for subject $i$, and thus account for subject-specific effects and dependencies among repeated (functional) observations. The replicate-specific random effects $\{\tilde \omega_{i,j}(\cdot)\}$ account for within-curve dependencies (or ``smooth" errors) that are unexplained by the fixed effects or subject-specific random effects. These effects are especially important for functional regression, even without repeated measurements
\citep{reiss_fast_2010,kowal_bayesian_2020}. Finally, the  errors $\{\epsilon_{i,j}(\cdot)\}$ describe any (non-smooth) measurement errors that remain. The random effects functions and errors $\{\tilde \gamma_i(\cdot), \tilde \omega_{i,j}(\cdot), \epsilon_{i,j}(\cdot)\}$ are mean zero and mutually independent, with additional modeling assumptions discussed subsequently.

The FMM \eqref{fmm} is widely applicable and has appeared in many previous studies \citep{zipunnikov_longitudinal_2014, cederbaum_functional_2016, zhu_fmem_2019,  li_fixedeffects_2022}, including both special cases without replications ($m_i =1$; \citealp{guo_functional_2002}) or covariates \citep{park_longitudinal_2015} as well as generalizations for additive \citep{scheipl_functional_2015} or non-Gaussian responses \citep{scheipl_generalized_2016}. However, computational considerations are paramount, and often preclude the use of these generalized models even for moderately-sized datasets \citep{sergazinov_case_2023}. In response, \citet{cui_fast_2022} proposed Fast Univariate Inference (FUI) for \eqref{fmm}, which seeks to dramatically simplify estimation by (i) fitting separate linear mixed models pointwise for each $\tau$, (ii) applying a smoother to the estimated pointwise fixed effects, and (iii) using asymptotic arguments or bootstrapping to obtain confidence bands for $\{\tilde \alpha_\ell(\cdot)\}$. Such a deconstruction sacrifices estimation efficiency and is difficult to apply for sparsely- or irregularly-sampled functional data. Like other frequentist approaches, FUI requires selection of tuning (smoothing) parameters and provides limited uncertainty quantification for certain parameters and predictions. Yet most important, we show subsequently  that such a decomposition is \emph{not} necessary to achieve scalable computing---and in fact that FUI is slower than our fully Bayesian approach (see Section~\ref{sec:sims}).

In general, Bayesian approaches for FMMs are highly appealing due to the consolidated interpretation of fixed and random effects, as well as convenient uncertainty quantification of all model parameters and predictions. \citet{morris_wavelet-based_2006} introduced Bayesian FMMs using a wavelet basis expansion for fixed and random effects. This approach has been extended and applied broadly \citep{zhu_robust_2011,morris_comparison_2017, lee_bayesian_2019}, but leverages unique features of the wavelet basis---which may not be suitable for smoother functional data and is difficult to apply when the number of functional observations is non-dyadic---and requires Metropolis-Hastings sampling steps for all variance components. Bayesian FMMs commonly rely on Markov chain Monte Carlo (MCMC) sampling for posterior inference, typically using Gibbs sampling \citep{morris_wavelet-based_2006, goldsmith_assessing_2016, lee_bayesian_2019} or Hamiltonian Monte Carlo \citep{goldsmith_generalized_2015}. Advantageously, these algorithms provide exact (up to Monte Carlo error) inference, but are prohibitively slow, with running times in the hours or days even for moderately-sized datasets (see Section~\ref{sec:sims}). Further, certain MCMC sampling strategies are vulnerable to slow mixing and convergence: for example, \citet{goldsmith_assessing_2016} applied a Gibbs sampler that alternates sampling blocks for the fixed effects and the random effects, which is sensitive to the model parameterization (i.e., centered vs. noncentered; \citealp{yu_center_2011}) and can lead to poor exploration of the joint target distribution. Our analyses confirm the MCMC inefficiencies of this blocking strategy, which compounds the impact of the extremely lengthy  computing times. 



To reduce this computational burden, there has been recent development for variational Bayes (VB) approximations for Bayesian FMMs. VB substantially reduces computation times compared to existing MCMC algorithms
\citep{goldsmith_assessing_2016, huo_ultra-fast_2022}. However, VB often provides poor uncertainty quantification, and thus \citet{goldsmith_assessing_2016} recommend it primarily as a tool to obtain quick initial estimates for model-building. Our simulations (Section~\ref{sec:sims}) show that VB is adequate for point estimation, but falls considerably short in interval estimation. Thus, there is urgent demand for algorithms that can provide both \emph{accurate} and \emph{scalable} Bayesian inference for the FMM \eqref{fmm}. 


We address this significant gap in the literature. Specifically, we design an MCMC sampling algorithm for the FMM \eqref{fmm} that offers several unique features. First, we jointly sample \emph{all} fixed and random effects functions $\{\tilde{\alpha}_{\ell}(\cdot), \tilde{\gamma}_i(\cdot), \tilde{\omega}_{i,j}(\cdot)  \}$, which delivers superior Monte Carlo efficiency for these critical quantities. 
Second, we design this joint sampler using a careful blocking structure paired with an orthogonalized basis reparametrization, which leads to exceptionally fast sampling steps. The accompanying variance components are sampled in a  separate block, yielding a convenient Gibbs sampler that is both \emph{computationally} and \emph{Monte Carlo} efficient. Perhaps surprisingly, our new MCMC sampling algorithm even surpasses state-of-the-art algorithms for frequentist estimation \citep{cui_fast_2022, li_fixedeffects_2022}
and VB approximations \citep{goldsmith_assessing_2016} and is orders of magnitude faster than existing Gibbs samplers for FMMs. This superior scaling persists across number of subjects $n$, number of replicates $m_i$ per subject, and number of covariates $L$, and is accompanied by more accurate point estimation and uncertainty quantification across nearly all tested scenarios. Applying our methods to the NHANES PA dataset, we demonstrate the significant practical impacts of our improved MCMC algorithm and provide posterior uncertainty quantification for key parameters. Compared to existing MCMC algorithms, we reduce the computation time from two weeks to only a few minutes. 

The rest of this article is organized as follows. Section \ref{sec:model} presents the Bayesian FMM for longitudinal data. Section \ref{sec:mcmc} describes the MCMC algorithm. Section \ref{sec:sims} provides simulation analyses. We apply our model on the NHANES dataset in Section \ref{sec:app}. We conclude with a discussion in Section \ref{sec:disc}. \texttt{R} code to implement our approach and replicate our results is available at \texttt{\url{https://github.com/thomasysun/FLFOSR}}.

\section{Basis expansions and prior distributions}
\label{sec:model}

FMMs \eqref{fmm} are most commonly implemented using basis expansions, including 
splines \citep{guo_functional_2002,goldsmith_assessing_2016},
functional principal components \citep{park_longitudinal_2015,li_fixedeffects_2022},
 and wavelets \citep{morris_wavelet-based_2006,huo_ultra-fast_2022}. The general basis representation of \eqref{fmm} is 
\begin{align}
    \label{basis-1}
    Y_{i,j}(\tau) &= \sum_{k=1}^{K} b_k(\tau)\beta_{k,i,j} + \epsilon_{i,j}(\tau), \quad \epsilon_{i,j}\left(\tau\right) \overset{iid}{\sim} N(0, \sigma^2_{\epsilon}) \\
    \label{basis-2}
    \beta_{k,i,j} &= \alpha_{k,0} + \sum_{\ell=1}^{L} x_{i,j,\ell}\alpha_{k,\ell} + \gamma_{k,i} + \omega_{k,i,j}
\end{align}
where $\{b_k(\cdot)\}_{k=1}^K$ are known basis functions and $\{\alpha_{k,\ell}, \gamma_{k,i}, \omega_{k,i,j}\}$ are unknown basis coefficients. Model \eqref{basis-1}--\eqref{basis-2} induces \eqref{fmm} under the identification $\tilde \alpha_0(\tau) = \sum_{k=1}^K b_k(\tau) \alpha_{k,0}$ and similarly for the remaining fixed and random effects functions. The coefficients $\{\beta_{k,i,j}\}$ are completely determined by $\{\alpha_{k,\ell}, \gamma_{k,i}, \omega_{k,i,j}\}$, and merely serve as a placeholder for notational convenience. 
Because the basis functions $\{b_k\}$ are known, estimation and inference for the coefficients $\{\alpha_{k,\ell}, \gamma_{k,i}, \omega_{k,i,j}\}$ is sufficient for estimation and inference of the \emph{functions} $\{\tilde \alpha_{\ell}(\cdot), \tilde\gamma_{i}(\cdot), \tilde \omega_{i,j}(\cdot)\}$. 

The choice of basis functions $\{b_k\}$ must be paired carefully with the choice of prior on the coefficients $\{\alpha_{k,\ell}, \gamma_{k,i}, \omega_{k,i,j}\}$, which together induce a prior  for functions $\{\tilde \alpha_{\ell}(\cdot), \gamma_{i}(\cdot), \tilde \omega_{i,j}(\cdot)\}$. Typically, we select the prior on the coefficients to encourage certain properties for the functions,  such as smoothness \citep{goldsmith_generalized_2015,goldsmith_assessing_2016,kowal_bayesian_2020} or sparsity \citep{morris_wavelet-based_2006}. Here, we also prioritize the resulting computational implications.

To motivate the general approach, suppose that we are interested in specifying a prior for a generic function under a basis expansion:
\[
\tilde \zeta(\tau) = \sum_{k=1}^K b_k(\tau) \zeta_k.
\]
A common strategy is to assume a prior of the form 
\[
\bm \zeta = (\zeta_1,\ldots, \zeta_K)' \sim N_K(\bm 0, \sigma_\zeta^2 \bm P^{-}),
\]
where $\bm P$ is a known (roughness) penalty matrix and $\sigma_\zeta^2$ is a (smoothness) parameter. For instance, suppose $\bm P$ is the matrix of integrated squared second derivatives, $[\bm P]_{k,k'} = \int {b}_k''(\tau) {b}_{k'}''(\tau) d\tau$, where $ b_k''(\cdot)$ denotes the second derivative of $b_k(\cdot)$. Then the prior for $\bm \zeta$  may be expressed as
\[
-2\log p(\bm \zeta \mid \sigma_\zeta^2) \stackrel{c}{=} \sigma_\zeta^{-2} \bm \zeta' \bm P \bm \zeta = \sigma_\zeta^{-2} \int \Big\{ \tilde \zeta''(\tau)\Big\}^2 d\tau
\]
and $\stackrel{c}{=}$ denotes equality up to an additive constant. Thus, the log-prior on $\bm \zeta$ reproduces the classical roughness penalty on the function $\tilde \zeta(\cdot)$. 
In a Bayesian framework, we further may place a prior on $\sigma_\zeta^2$ to learn the smoothness parameter; these details are discussed below in the context of model \eqref{basis-1}--\eqref{basis-2}.  

Although this approach is advantageous due to its generality and smoothness-inducing properties, it is unfavorable for computing, especially within the basis-expanded FMM \eqref{basis-1}--\eqref{basis-2}. Suppose we observe data at  $\tau_1,\ldots,\tau_T \in \mathcal{T}$ and let $\boldsymbol{\mathcal{B}}_0 = (\bm b_1,\ldots, \bm b _K)$ be the $T \times K$ basis matrix with $\bm b_k = (b_k(\tau_1),\ldots, b_k(\tau_T))'$. For a generic basis $\{b_k\}$, there is no special structure for $\boldsymbol{\mathcal{B}}_0$ or  $\bm P^-$. Thus, we seek to reparamaterize these terms $(\boldsymbol{\mathcal{B}}_0, \bm P)$---at a one-time cost for all subsequent MCMC sampling---for more amenable computing. Specifically, we apply the orthogonalization strategy from \citet{scheipl_spike-and-slab_2012}, which uses the spectral decomposition of $\boldsymbol{\mathcal{B}}_0 \bm P^- \boldsymbol{\mathcal{B}}_0'$ to form a basis matrix $\boldsymbol{\mathcal{B}}$ such that (i) $\boldsymbol{\mathcal{B}}' \boldsymbol{\mathcal{B}} = \mbox{diag}(\{d_k\}^{K}_{k=1})$ is a diagonal matrix and (ii) $ (\tilde \zeta(\tau_1),\ldots, \tilde \zeta(\tau_T))'  = \boldsymbol{\mathcal{B}}_0\bm \zeta $ has the same distribution as $ \boldsymbol{\mathcal{B}} \bm \zeta^*$, where $\bm \zeta^* \sim N(\bm 0, \sigma_\zeta^2 \bm I)$. Notably, this strategy applies for any basis $\{b_k\}$ and penalty matrix $\bm P^-$; in our empirical examples, we use B-splines with a penalty on the second differences of the coefficients. 


Revisiting the FMM basis expansion \eqref{basis-1}--\eqref{basis-2}, we henceforth assume, without loss of generality, that $\boldsymbol{\mathcal{B}}' \boldsymbol{\mathcal{B}}$ is diagonal and that arbitrary (roughness) penalties may be incorporated via independent Gaussian priors. Thus, we  
specify the priors
\begin{equation}
    \label{priors}
    [\alpha_{k,\ell} \mid \sigma_{\alpha_{\ell}}^2] \stackrel{indep}{\sim} N(0, \sigma_{\alpha_{\ell}}^2), \quad    
    [\gamma_{k,i} \mid \sigma_\gamma^2] \stackrel{indep}{\sim} N(0, \sigma_\gamma^2),
    \quad 
        [\omega_{k,i,j} \mid \sigma_{\omega_i}^2] \stackrel{indep}{\sim} N(0, \sigma_{\omega_i}^2)
\end{equation}
and assume that the reparametrization has already been completed. In conjunction with the basis expansions,  the priors \eqref{priors} induce a Gaussian process prior for each fixed and random effect function, e.g., 
\[
\tilde \alpha_0 \sim \mathcal{GP}(0, C_{\alpha_0}), \quad C_{\alpha_0}(\tau, u)=\sigma_{\alpha_{0}}^2\sum_{k=1}^K b_k(\tau) b_k(u) 
\]
and similarly for the remaining terms.

Finally, we assume conditionally conjugate Gamma priors for the precision parameters,
\begin{equation}
    \label{priors-vc}
\{\sigma_{\alpha_\ell}^{-2}, \sigma_{\gamma}^{-2}, \sigma_{\omega_i}^{-2}\} \stackrel{iid}{\sim} \mbox{Gamma}(a,b)
\end{equation}
along with a Jeffreys' prior for the observation error variance,  $[\sigma^2_{\epsilon}] \propto 1/\sigma^2_{\epsilon}$. These parameters determine both the smoothness of the corresponding functions as well as the  various sources of variability within a function, between replicates for a subject, and between subjects.  
Other priors are  available for variance parameters \citep{gelman_prior_2006}, and may be substituted into the proposed framework with minimal impact on the core MCMC sampling approach. However, we find that \eqref{priors-vc} offers excellent modeling performance and MCMC efficiency, and  is not highly sensitive to the choice of $(a,b)$ (see the supplementary material for a sensitivity analysis).

In practice, we observe the functional data $Y_{i,j}(\tau)$ at discrete points $\tau_1,\ldots,\tau_T \in \mathcal{T}$. For simplicity, we assume that these observation points are common for all $(i,j)$, but it is straightforward to accommodate sparsely- or irregularly-sampled functional data within a Bayesian framework \citep{kowal_functional_2019}. The likelihood under \eqref{basis-1} with observed data $\bm Y_{i,j} = (Y_{i,j}(\tau_1),\ldots, Y_{i,j}(\tau_T))'$ is 
\begin{equation}
    \label{like}
    \bm Y_{i,j} = \boldsymbol{\mathcal{B}} \bm \beta_{i,j} + \bm\epsilon_{i,j}, \quad \bm \epsilon_{i,j} \stackrel{iid}{\sim} N_T(\bm 0, \sigma_\epsilon^2 \bm I_T)
\end{equation}
where $\bm \beta_{i,j} = (\beta_{i,j,1},\ldots, \beta_{i,j,K})'$ is modeled via \eqref{basis-2}--\eqref{priors}. 

We emphasize that the modeling choices in \eqref{basis-1}--\eqref{priors-vc} are meant to be broadly applicable---including generic basis expansions, penalty matrices, and variance components (or smoothness parameters) for each  fixed and random effects function. The main contributions of this paper are not found in the uniqueness of this modeling strategy, but rather in our  MCMC sampling algorithm for the general Bayesian FMM \eqref{fmm}, which is presented in the next section. The essential features of our FMM  specification  are  (i) the basis expansions \eqref{basis-1} and (ii) the conditionally Gaussian priors \eqref{priors},  in conjunction with the aforementioned basis orthogonalization strategy \citep{scheipl_spike-and-slab_2012}.

\section{MCMC Algorithm}\label{sec:mcmc}
The primary challenge is to construct a sampling algorithm for the joint posterior distribution of the fixed and random effects functions $\{\tilde{\alpha}_{\ell}(\cdot), \tilde{\gamma}_i(\cdot), \tilde{\omega}_{i,j}(\cdot)  \}$ under the FMM \eqref{fmm}. This task requires simultaneous consideration of (i) MCMC efficiency, including convergence and autocorrelation of the Markov chain, and (ii) computational scalability across all dimensions of the data: the number of subjects $n$, the number of replicates $m_i$ per subject, number of covariates $L$, and the number of observation points $T$ along each curve. We focus on the sampling steps for the fixed and random effects basis coefficients   $\{\alpha_{k,\ell}, \gamma_{k,i}, \omega_{k,i,j}\}$, which are sufficient for posterior inference for the fixed and random effects functions  $\{\tilde \alpha_{\ell}(\cdot), \tilde\gamma_{i}(\cdot), \tilde \omega_{i,j}(\cdot)\}$ under \eqref{basis-1}--\eqref{basis-2}. The variance components $\{\sigma_{\alpha_\ell}^{2}, \sigma_{\gamma}^{2}, \sigma_{\omega_i}^{2}\}$ are sampled  in a separate block, resulting in a two-block Gibbs sampling algorithm (Algorithm~\ref{alg:mcmc}).

We propose a \emph{joint} sampler for \emph{all} fixed and random effects functions $\{\tilde \alpha_{\ell}(\cdot), \tilde\gamma_{i}(\cdot), \tilde \omega_{i,j}(\cdot)\}$ by carefully decomposing the joint posterior distribution of the corresponding basis coefficients. First, let $\bm{\alpha}_k = (\alpha_{k,0}, \alpha_{k,1}, \dots, \alpha_{k, \ell})'$, $\bm{\gamma}_k = (\gamma_{k,1}, \dots, \gamma_{k,n})'$, and $\bm{\omega}_k = (\bm{\omega}_{k,1}', \dots, \bm{\omega}_{k,n}')'$, where $\bm{\omega}_{k, i} =  (\omega_{k, i, 1}, \dots, \omega_{k, i, m_i})'$, and  let $\boldsymbol{\alpha} = (\boldsymbol{\alpha}_1, \dots, \boldsymbol{\alpha}_K)^{\prime}$ and similarly for $\boldsymbol{\gamma}$ and $\boldsymbol{\omega}$. Similarly, define the diagonal variance matrices $ \boldsymbol{\Sigma}_\epsilon = \sigma^2_{\epsilon}\textbf{I}_{M}$, $ \boldsymbol{\Sigma}_\alpha = \operatorname{diag}\left(\{\sigma_{\alpha_{\ell}}^2 \}_{\ell=1}^{L}\right)$, $ \boldsymbol{\Sigma}_\gamma = \operatorname{diag}\left(\{\sigma_{\gamma_{}}^2 \}_{i=1}^{n}\right) = \sigma^2_{\gamma}\textbf{I}_{n}$, and $\boldsymbol{\Sigma}_\omega = \operatorname{diag}\left(\{\sigma_{\omega_{i}}^2 \}_{i,j}\right)$, where the set of variances are first iterated through $j = 1, \dots, m_i$ for $i=1$ and so on to yield $M$ terms along the diagonal. Let $\boldsymbol{\Sigma} = \{\boldsymbol{\Sigma}_\alpha, \boldsymbol{\Sigma}_\gamma ,  \boldsymbol{\Sigma}_\omega, \boldsymbol{\Sigma}_\epsilon\}$ contain all variance components.

The joint fixed and random effects posterior (conditional on $\bm \Sigma$) is  decomposable as 
\begin{equation}
    \label{joint-decomp}
    p(\bm \alpha, \bm \gamma, \bm \omega \mid \bm Y, \bm \Sigma) = p(\bm \alpha \mid \bm Y, \bm \Sigma)\  p(\bm \gamma \mid \bm Y,  \bm \alpha, \bm \Sigma )\  p(\bm \omega \mid \bm Y,  \bm \alpha, \bm \gamma, \bm \Sigma)
\end{equation}
where $\bm Y = \{\bm Y_{i,j}\}$ denotes all observed data. Our Gibbs sampler iterates between joint sampling blocks from \eqref{joint-decomp} and $[\bm \Sigma \mid \bm Y, \bm \alpha, \bm \gamma, \bm \omega ]$; the latter sampling step is straightforward (see Algorithm~\ref{alg:mcmc}), so  we focus on \eqref{joint-decomp}. 

We  sample from the joint posterior \eqref{joint-decomp} by iteratively drawing from the three constituent terms: (i) the marginal posterior of the fixed effects $\bm \alpha$, (ii) the partial conditional posterior of the subject-specific random effects $\bm \gamma$ given the fixed effects $\bm \alpha$, and (iii) the full conditional posterior of the replicate-specific random effects $\bm \omega$. In contrast to a Gibbs sampler that draws from the three \emph{full conditional} distributions $[\bm \alpha \mid \bm Y, \bm \gamma, \bm \omega, \bm \Sigma]$, $[\bm \gamma \mid \bm Y, \bm \alpha, \bm \omega, \bm \Sigma]$, and $[\bm \omega \mid \bm Y, \bm \alpha, \bm \gamma, \bm \Sigma]$ (e.g., \citealp{goldsmith_assessing_2016}), our approach delivers a direct Monte Carlo (not MCMC) sample from the joint posterior (conditional on $\bm \Sigma$), and thus offers the potential for large gains in sampling efficiency (see Figure~\ref{fig:mcmceff}). Further, the joint sampler requires no consideration of centered versus non-centered parameterizations of the mixed effects model, which eliminates a recurring nuisance for Bayesian implementations of mixed and hierarchical models \citep{yu_center_2011}. However, we must also  carefully consider the resulting computational burden of these sampling steps. In particular, a sampling algorithm that offers Monte Carlo efficiency may nonetheless be infeasible in practice, if the raw computing times scale poorly in $n$, $m_i$, $L$, or $T$. 

Our key innovation is that we provide convenient, closed-form, and highly scalable sampling steps for the  constituent  distributions in \eqref{joint-decomp}. First, observe that $\{\beta_{k,i,j}\}$ in \eqref{basis-1}--\eqref{basis-2} contains all fixed and random effects coefficients, and thus may be viewed as a placeholder for each of the random variables in \eqref{joint-decomp}. Since the reparametrized basis matrix  satisfies  $\boldsymbol{\mathcal{B}}' \boldsymbol{\mathcal{B}} = \mbox{diag}(\{d_k\}_{k=1}^K)$, the joint likelihood \eqref{like} for $\{\beta_{k,i,j}\}$ may be written
\begin{align}
    \label{like-1}
    p(\bm  Y \mid \{\beta_{k,i,j}\}, \bm \Sigma) &\propto 
    \prod_{i=1}^n \prod_{j=1}^{m_i} \exp \left\{-\frac{1}{2 \sigma_\epsilon^2} \left\Vert \bm{Y}_{i,j}-\boldsymbol{\mathcal{B}}\bm{\beta}_{i,j} \right\Vert_2^2 \right\} \\ 
    \label{like-2}
    &\propto \exp \left\{-\frac{1}{2 \sigma_\epsilon^2} \sum_{i=1}^n \sum_{j=1}^{m_i} \sum_{k=1}^K d_k (y_{k,i,j}  - \beta_{k,i,j})^2
    \right\}
\end{align}
up to constants that do not depend on $\{\beta_{k,i,j}\}$, 
where the coefficients $\{y_{k,i,j}\}$ are computed by projecting the functional observations onto the basis matrix, $(\boldsymbol{\mathcal{B}}'\boldsymbol{\mathcal{B}})^{-1}\boldsymbol{\mathcal{B}}'\bm Y_{i,j} = (y_{1,i,j}, \ldots, y_{K, i, j})$. Thus, the only dependence on $T$ is via this projection step, which is a one-time cost for all subsequent MCMC sampling. 
In addition, 
the joint posterior factorizes across the basis coefficients, 
\[
p(\bm \alpha, \bm \gamma, \bm \omega \mid \bm Y, \bm \Sigma) = \prod_{k=1}^K p(\bm \alpha_k, \bm \gamma_k, \bm \omega_k \mid \bm Y, \bm \Sigma),
\]
and  we may sample the fixed and random effects coefficients separately for each $k=1,\ldots, K$. This strategy is parallelizable across $k$ yet  still maintains a joint sampler for all fixed and random effects coefficients. This offers a substantial simplification from the 
functional data likelihood in \eqref{like-1}, which features $T$-dimensional terms $\bm Y_{i,j}$, especially since $K \ll T$ in general. Finally, the likelihood \eqref{like-2} for  $\{\beta_{k,i,j}\}$ is proportional to the likelihood for $\{\beta_{k,i,j}\}$ implied by the simple \emph{working model}
\begin{align}
\label{like-work-1}
y_{k,i,j} &= \beta_{k,i,j} + \epsilon_{k,i,j}, \quad \epsilon_{k,i,j}\stackrel{indep}{\sim}N(0, \sigma_\epsilon^2/d_k) \\
\label{like-work-2}
&= \alpha_{k,0} + \sum_{\ell=1}^{L} x_{i,j,\ell}\alpha_{k,\ell} + \gamma_{k,i} + \omega_{k,i,j} + \epsilon_{k,i,j}.
\end{align}
It is typically easier to derive the posterior distributions for  $\{\alpha_{k,\ell}, \gamma_{k,i}, \omega_{k,i,j}\}$ in \eqref{joint-decomp} using \eqref{like-work-1}--\eqref{like-work-2} instead of \eqref{like} or \eqref{like-1}, yet the results are equivalent. 

We describe the three sampling steps in ascending complexity, which reverses the actual sampling order from \eqref{joint-decomp}. Let $\bm y_k = (y_{k,1,1}, \ldots, y_{k,1, m_1}, \ldots, y_{k, n, m_n})'$ denote the projected functional data ordered by replicates for each subject, 
$\boldsymbol{X}$ be the $M \times (L+1)$ design matrix with a column of ones for the intercept,  and $\textbf{Z}=\operatorname{bdiag}\left\{\mathbf{1}_{m_i}\right\}_{i=1}^n$, a block diagonal matrix with $n$ columns and $m_i$-dimensional vectors of ones. 

The simplest sampling step is the full conditional distribution for the replicate-specific coefficients: 
\begin{equation}
\label{post-omega}
\begin{gathered}
[\boldsymbol{\omega}_k \mid \boldsymbol{Y},\boldsymbol{\alpha} , \boldsymbol{\gamma},  \boldsymbol{\Sigma}] \stackrel{indep}{\sim} N_M(\boldsymbol{Q}^{-1}_{\omega_k} \boldsymbol{\ell}_{\omega_k},  \boldsymbol{Q}^{-1}_{\omega_k}) \\
   \boldsymbol{Q}_{\omega_k} = \operatorname{diag}\left( \{d_k\sigma_{\epsilon}^{-2} + \sigma_{\omega_{i}}^{-2} \}_{i,j}\right) \\ 
   \boldsymbol{\ell}_{\omega_k} = d_k\sigma_{\epsilon}^{-2} \left\{\boldsymbol{y}_k - \left(\boldsymbol{X}\boldsymbol{\alpha}_k + \textbf{Z}\boldsymbol{\gamma}_k \right) \right\}
\end{gathered}
\end{equation}
for $k=1,\ldots,K$. 
Most important, the posterior precision matrix $\bm Q_{\omega_k}$ is diagonal, and thus all the replicate-specific coefficients $\bm \omega$ may be sampled independently (and in parallel) in $\mathcal{O}(MK)$ computing time. 

Next, we provide the sampling step for the partial conditional of the subject-specific random effects, $[\bm \gamma_k \mid \bm Y, \bm \alpha, \bm \Sigma]$, which marginalizes over $\bm \omega$. Under the prior \eqref{priors}, it is easy to see that the marginalized version of \eqref{like-work-2} is simply 
$y_{k,i,j} = \alpha_{k,0} + \sum_{\ell=1}^{L} x_{i,j,\ell}\alpha_{k,\ell} + \gamma_{k,i} + \nu_{k,i,j}$, where $\nu_{k,i,j} = \omega_{k,i,j} + \epsilon_{k,i,j} \stackrel{indep}{\sim} N(0, \sigma_{\omega_i}^2 + \sigma_\epsilon^2/d_k)$. The requisite distribution is then
\begin{equation}
    \label{post-gamma}
    \begin{gathered}
    [\bm \gamma_k \mid \bm Y, \bm \alpha, \bm \Sigma] \stackrel{indep}{\sim} N_n(\boldsymbol{Q}^{-1}_{\gamma_k} \boldsymbol{\ell}_{\gamma_k},  \boldsymbol{Q}^{-1}_{\gamma_k}) \\
    \boldsymbol{Q}_{\gamma_k} = \boldsymbol{\Sigma}_\gamma^{-1} + \textbf{Z}^{\prime} \boldsymbol{V}_k\textbf{Z}^{} = \operatorname{diag}\left(\{(\sigma_{\gamma}^{-2} + d_k m_i(\sigma_{\epsilon}^2 + d_k\sigma_{\omega_{i}}^2)^{-1}) \}_{i=1}^{n}\right)\\
    \boldsymbol{\ell}_{\gamma_k} = d_k\textbf{Z}^{\prime} \boldsymbol{V}_k \left(\boldsymbol{y}_k - \boldsymbol{X}\boldsymbol{\alpha}_k \right)
    \end{gathered}
\end{equation}
where $\boldsymbol{V}_k =  \boldsymbol{\Sigma}_{\epsilon}^{-1} - d_k\boldsymbol{\Sigma}_{\epsilon}^{-1} \left( d_k\boldsymbol{\Sigma}_{\omega}^{-1} + \boldsymbol{\Sigma}_{\epsilon}^{-1}\right)^{-1} \boldsymbol{\Sigma}_{\epsilon}^{-1} = \operatorname{diag}\left(\{(\sigma_{\epsilon}^2 + d_k\sigma_{\omega_{i}}^2)^{-1} \}_{i,j}\right)$ for $k=1,\ldots,K$. Crucially, the posterior precision matrix $\bm Q_{\gamma_k}$ is diagonal, and thus all subject-specific coefficients $\bm \gamma$ may be sampled independently (and in parallel) in $\mathcal{O}(nK)$ computing time. 

While the posterior precision matrices for $[\bm \omega_k \mid \bm Y, \bm \alpha, \bm \gamma, \bm \Sigma]$ and $[\bm \gamma_k \mid \bm Y, \bm \alpha, \bm \Sigma]$ are diagonal and easy to construct, we must also consider computation of the vectors $\boldsymbol{\ell}_{\omega_k}$ and $\boldsymbol{\ell}_{\gamma_k}$. An important observation is that, for arbitrary matrix $\Theta_{n\times n}$, the operations $\textbf{Z}\Theta_{n\times n}$  (or $\Theta_{n\times n} \textbf{Z}^{\prime}$) repeat row $i$ (or column $i$) of $\Theta_{n\times n}$ $m_i$ times, while $\textbf{Z}^{\prime}\Theta_{M\times M}$ (or $\Theta_{M\times M} \textbf{Z}$) sums over the columns (or rows) for all rows (or columns) corresponding to the same subject $i$. The matrix notation obscures the simplicity of these computations, which are vectorizable and thus offer additional efficiency gains in \texttt{R}. 


The remaining sampling step is the marginal posterior of the fixed effects, $[\bm \alpha_k \mid \bm Y, \bm \Sigma]$, which requires marginalization over \emph{all} random effects coefficients $\{\bm \gamma, \bm \omega\}$. Once more, the working model \eqref{like-work-1}--\eqref{like-work-2} is particularly convenient for these derivations, and yields the distribution
\begin{equation}   
\label{post-alpha}
\begin{gathered}
    [\boldsymbol{\alpha}_k \mid \boldsymbol{Y}, \boldsymbol{\Sigma}] \stackrel{indep}{\sim} N_L(\boldsymbol{Q}^{-1}_{\alpha_k} \boldsymbol{\ell}_{\alpha_k},  \boldsymbol{Q}^{-1}_{\alpha_k}) \\
    \boldsymbol{Q}_{\alpha_k} = \boldsymbol{\Sigma}_\alpha^{-1} + d_k\boldsymbol{X}^{\prime} \left(\boldsymbol{V}_k - \boldsymbol{W}_k \right)\boldsymbol{X}\\ \boldsymbol{\ell}_{\alpha_k} = d_k \boldsymbol{X} ^{\prime} \left(\boldsymbol{V}_k - \boldsymbol{W}_k \right)\boldsymbol{y}_k
\end{gathered}
\end{equation}
where $\bm W_k = (\textbf{Z}^{\prime} \boldsymbol{V}_k)^{\prime} ( \boldsymbol{\Sigma}_\gamma^{-1} + d_k\textbf{Z}^{\prime}\boldsymbol{V}_k\textbf{Z})^{-1}\textbf{Z}^{\prime} \boldsymbol{V}_k$ and  $\boldsymbol{V}_k - \boldsymbol{W}_k = \operatorname{diag}\big(\{(\sigma_{\epsilon}^2 + d_k\sigma_{\omega_{i}}^2 + \\ d_k m_i \sigma_{\gamma}^2)^{-1} \}_{i,j}\big)$ for $k=1,\ldots, K$. Unlike for the random effects coefficients, the posterior precision for $\bm \alpha_k$ is not diagonal; indeed, $\bm Q_{\alpha_k}$ is the \emph{only} non-diagonal posterior precision matrix in our joint sampler for all $\{\alpha_{k,\ell}, \gamma_{k,i}, \omega_{k,i,j}\}$. To alleviate this potential bottleneck when the number of covariates $L$ is large, we apply the $\mathcal{O}(L^3)$ sampling algorithm from \citet{rue_fast_2001} when $L \leq M$ and the $\mathcal{O}(M^2 L)$ sampling algorithm from \citet{bhattacharya_fast_2016} when $L > M$. This sampling strategy has been successful in (non-longitudinal) function-on-scalars regression \citep{kowal_bayesian_2020}. More generally, \eqref{post-alpha} has the same structure as in (non-functional) Bayesian linear regression models, and thus we may adapt and apply new sampling strategies for that setting as the state-of-the-art advances \citep{nishimura_prior-preconditioned_2022}.

In aggregate, we present our full MCMC sampling algorithm in Algorithm~\ref{alg:mcmc}. The joint sampling step for $[\bm \alpha, \bm \gamma, \bm \omega \mid \bm Y, \bm \Sigma]$ is highly scalable, and in fact delivers the same computational cost as a naive Gibbs sampler that instead use the three full conditional draws  $[\bm \alpha \mid \bm Y, \bm \gamma, \bm \omega, \bm \Sigma]$, $[\bm \gamma \mid \bm Y, \bm \alpha, \bm \omega, \bm \Sigma]$, and $[\bm \omega \mid \bm Y, \bm \alpha, \bm \gamma, \bm \Sigma]$---while offering the potential for substantial increases in MCMC efficiency.

\begin{algorithm}
\SetAlgoLined
\begin{enumerate}
    \item Sample the fixed and random effects coefficients $[\bm \alpha, \bm \gamma, \bm \omega \mid \bm Y, \bm \Sigma]$:
    \begin{enumerate}
        \item Sample $[\bm \alpha \mid \bm Y, \bm \Sigma]$ from \eqref{post-alpha}
        \item Sample $[\bm \gamma \mid \bm Y, \bm \alpha, \bm \Sigma]$ from \eqref{post-gamma}
        \item Sample $[\bm \omega \mid \bm Y, \bm \alpha, \bm \gamma, \bm \Sigma]$ from \eqref{post-omega}
    \end{enumerate}
    and compute functions $\tilde \alpha_\ell(\cdot) = \sum_{k=1}^K b_k(\cdot) \alpha_{k,\ell}$ and similarly for all $\{\tilde \gamma_i(\cdot), \tilde \omega_{i,j}(\cdot)\}$.

    \item Sample the variance components $[\bm \Sigma \mid \bm Y, \bm \alpha, \bm \gamma, \bm \omega]$:
    \begin{enumerate}
        \item Sample $[\sigma^{-2}_{\epsilon}  \mid \bm Y, \bm \alpha, \bm \gamma, \bm \omega] \sim \operatorname{Gamma}\big(MT/2 ,  \sum^{n}_{i=1}\sum^{m_i}_{j=1}\left\Vert \bm{Y}_{i,j}-\boldsymbol{\mathcal{B}}\bm{\beta}_{i,j} \right\Vert_2^2/2\big) $
    
            \item Sample
    $[\sigma^{-2}_{\alpha_{\ell}} \mid \bm Y, \bm \alpha, \bm \gamma, \bm \omega] \sim \operatorname{Gamma}\big(a + K/2, b +  \sum^{K}_{k=1}\alpha_{k, \ell}^2/2\big)$, \quad
    \quad $\ell=1,\ldots,L$ 
    
            \item Sample
    $[\sigma^{-2}_\gamma\mid \bm Y, \bm \alpha, \bm \gamma, \bm \omega] \sim \operatorname{Gamma}\big(a + nK/2, b +  \sum^{K}_{k=1}\sum^{n}_{i=1}\gamma_{k, i}^2/2\big) $ 
    
\item Sample $[\sigma^{-2}_{\omega_{i}}\mid \bm Y, \bm \alpha, \bm \gamma, \bm \omega] \sim \operatorname{Gamma}\big(a + m_i K/2, b +   \sum^{K}_{k=1}\sum^{m_i}_{j=1}\omega_{k, i, j}^2/2\big)$, \quad $i=1,\ldots,n$.
    \end{enumerate}
\end{enumerate}
\caption{MCMC sampling algorithm for the Bayesian FMM \eqref{fmm}--\eqref{like}: \emph{fast longitudinal function-on-scalar regression} (FLFOSR).
\label{alg:mcmc}}
\end{algorithm}

From a practical perspective, we emphasize that many computations are parallelizable, including the sampling steps for all $\{\bm \gamma, \bm \omega\}$, while many constituent terms are highly vectorizable. 
For instance, $\boldsymbol{V}_k$  and $\boldsymbol{W}_k$ are diagonal, and computing $\left(\boldsymbol{V}_k - \boldsymbol{W}_k \right)\boldsymbol{X}$ simply consists of multiplying each row of the matrix $\boldsymbol{X}$ by the corresponding scalar value $(\sigma_{\epsilon}^2 + d_k\sigma_{\omega_{i}}^2 + d_k m_i \sigma_{\gamma}^2 )^{-1}$.
Within \texttt{R}, we maximize computational performance by avoiding matrix calculations wherever possible and opting for vectorized operations instead. This is in contrast to common FMM formulations that  utilize matrix operations with Kronecker products or large block matrices, and thus face significant computational bottlenecks.

Lastly, we reiterate that Algorithm~\ref{alg:mcmc} is applicable for \emph{any} Bayesian FMM that satisfies a basis expansion \eqref{basis-1}--\eqref{basis-2} and assumes (conditionally) Gaussian priors \eqref{priors}. Thus, our approach is broadly applicable for many choices of basis functions, and remains valid for generic smoothness priors 
using the basis orthogonalization strategy of   \citet{scheipl_spike-and-slab_2012}. We henceforth refer to our approach as \emph{fast longitudinal function-on-scalar regression} (FLFOSR).

\section{Simulations}
\label{sec:sims}
We conduct a series of simulation studies to evaluate the performance of our approach against  state-of-the-art Bayesian and frequentist competitors. Specifically, we assess (i) the MCMC efficiency and  computational costs and (ii) the estimation accuracy and uncertainty quantification under various simulation designs. 

\subsection{Simulation Design}
\label{sec:simdesign}
Functional observations are generated from the FMM \eqref{fmm} using the basis representation \eqref{basis-1}--\eqref{basis-2}. We fix $\alpha_{k,0}^*=1$ and 
simulate the coefficients $\alpha^*_{k, \ell} \sim N(0, \sigma_{\alpha_{}}^{2*})$, $\gamma^*_{k,i} \sim  N(0, \sigma_{\gamma_{}}^{2*})$, and $\omega^*_{k, i, j} \sim  N(0, \sigma_{\omega_{}}^{2*})$ independently, and then compute $\beta^*_{k,i,j} = \alpha^*_{k,0} + \sum_{\ell=1}^{L} x_{i,\ell}\alpha^*_{k,\ell} + \gamma^*_{k,i} + \omega^*_{k,i,j}$.  The variance components $\{\sigma_{\alpha_{}}^{2*}, \sigma^{2*}_{\gamma}, \sigma^{2*}_{\omega},\sigma^{2*}_{\epsilon}\}$ are varied in 
Sections~\ref{sec:simcomp}~and~\ref{sec:simacc} to evaluate performance under different sources of variability. Finally, functional data  $Y_{i,j}(\tau_t)$ are generated from \eqref{basis-1}, where the basis functions are orthogonalized B-spline basis functions and $\{\tau_t\}_{t=1}^T$ is a grid of $T = 144$ equally-spaced points in $[0,1]$, which emulates the PA data in Section~\ref{sec:app}. We emphasize that although this data-generating process resembles the proposed Bayesian FMM, the same model structure is shared among all competing methods, and thus does not unfairly favor our approach. 

\subsection{Competing methods}
\label{sec:comparison}
We compare the proposed FLFOSR approach against several Bayesian and frequentist methods for regression analysis with longitudinal functional data. The Bayesian competitors come from the widely-used \texttt{refund} package in \texttt{R} \citep{refund}, and feature both  the Gibbs sampler (\texttt{refund:Gibbs}) and the VB approximation (\texttt{refund:VB}) from \citet{goldsmith_assessing_2016}. Unlike FLFOSR, \texttt{refund:Gibbs} (i) uses the full conditional draws for all fixed and random effects instead of the joint sampler (Algorithm~\ref{alg:mcmc}), (ii) does not orthogonalize the basis functions, and (iii)  sets the hyperparameters of the variance components based on an initial estimate of the residual covariance matrix. We expect that (i) will inhibit MCMC efficiency and (ii) will decrease computational scalability. 

Among frequentist approaches, there is a rich collection of recent strategies and software. We focus on FUI \citep{cui_fast_2022}, which specifically advertises computational scalability for longitudinal function-on-scalar regression, and \citet{li_fixedeffects_2022},  
which provides fixed effects inference for a variety of longitudinal correlation structures. We choose to work with the exchangeable correlation model since this mirrors our model assumptions.  More general methods for FMMs or functional additive mixed models could be considered for this problem. However, the widely-used implementations (i.e., \texttt{pffr} in the \texttt{refund} package) are known to have severe bottlenecks for both computing and memory  \citep{cui_fast_2022, sergazinov_case_2023}. Thus, such methods were not included in the study. 

For FLFOSR, we set the hyperparameters to be $a = b = 0.1$ and assess sensitivity in the supplementary material. All methods used $K=15$ basis functions. Any remaining settings were fixed at the default choices in the provided \texttt{R} functions. Simulations were performed on a Windows desktop with a 3.60 GHz Intel Xeon CPU with 32 GB of RAM.

\subsection{Evaluating MCMC efficiency and computational scalability}\label{sec:simcomp}
Among Bayesian methods, we measure MCMC efficiency using effective sample sizes (ESS), which equivalently summarize the autocorrelation in the Markov chain  \citep{gelman_bayesian_2013}. For each MCMC sampler, we generate $N = 1000$ draws from a single chain after discarding the initial $N_{\text{burn}} = 1000$ draws as burn-in, and then compute the pointwise ESS  $N_{\text{eff}}\{\tilde{\alpha}_\ell\left(\tau_{t}\right)\}$ of each fixed effect regression coefficient function $\tilde \alpha_\ell(\cdot)$ for $\ell=1,\ldots,L$ and $t=1,\ldots,T$. We also record $s_N$ and  $s_{N_{\text{burn}}}$, the total running times (in seconds) to obtain $N$ and $N_{\text{burn}}$ draws, respectively. 

\begin{figure}[h]
\centering
\includegraphics[width=1\textwidth]{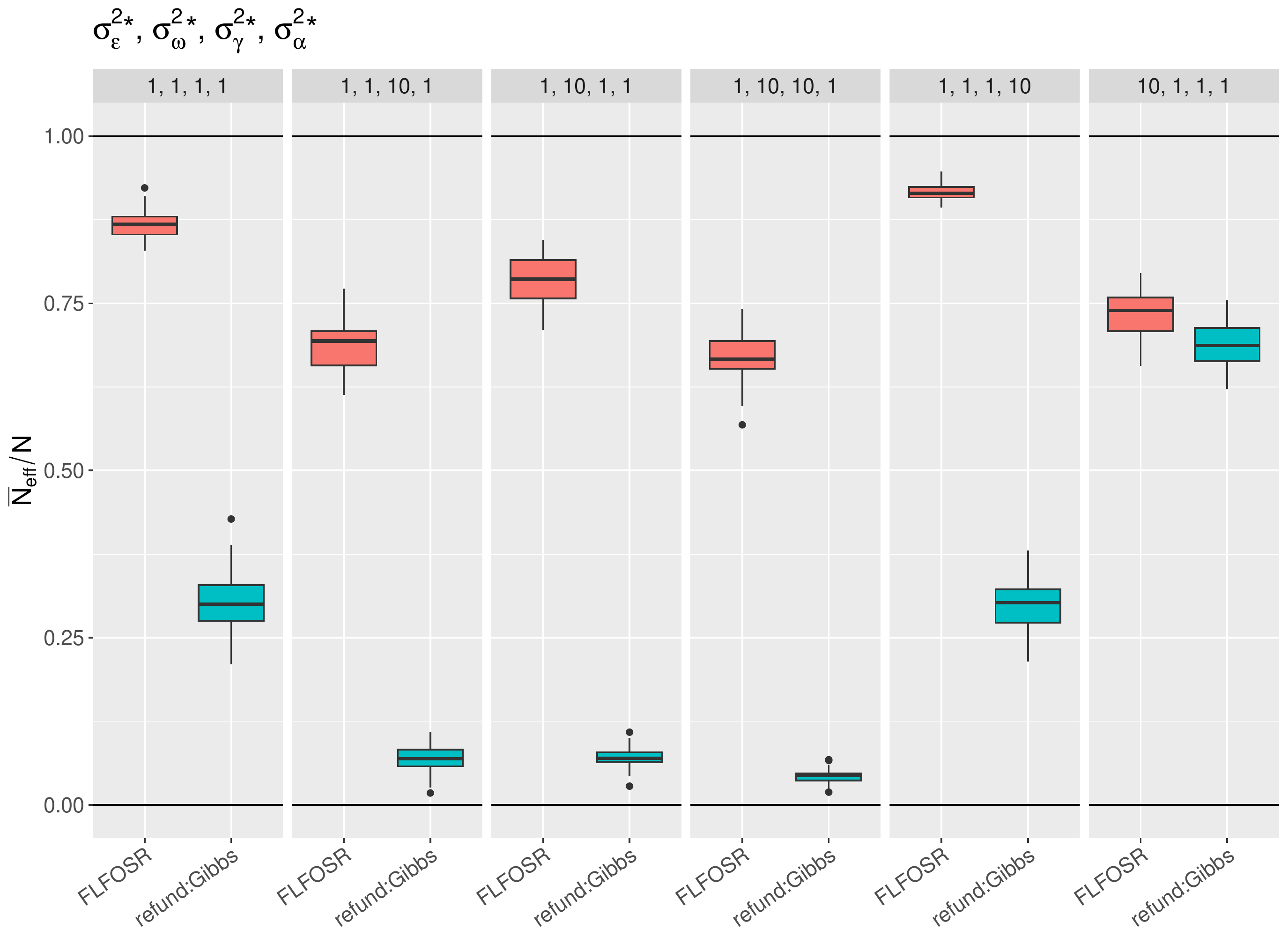}
\caption{Average relative efficiency $\Bar N_{\text{eff}}/N$ for FLFOSR and  \texttt{refund:Gibbs} across various simulation designs (columns) and 30 simulated datasets (boxplots). The proposed FLFOSR sampler shows consistently excellent and superior MCMC efficiency, while \texttt{refund:Gibbs} is extremely inefficient when
$\sigma^{2*}_{\omega}$ or $\sigma^{2*}_{\gamma}$ is large.}
  \label{fig:mcmceff}
\end{figure}

\begin{figure}
\begin{subfigure}{.5\linewidth}
\centering
\includegraphics[width=1\textwidth]{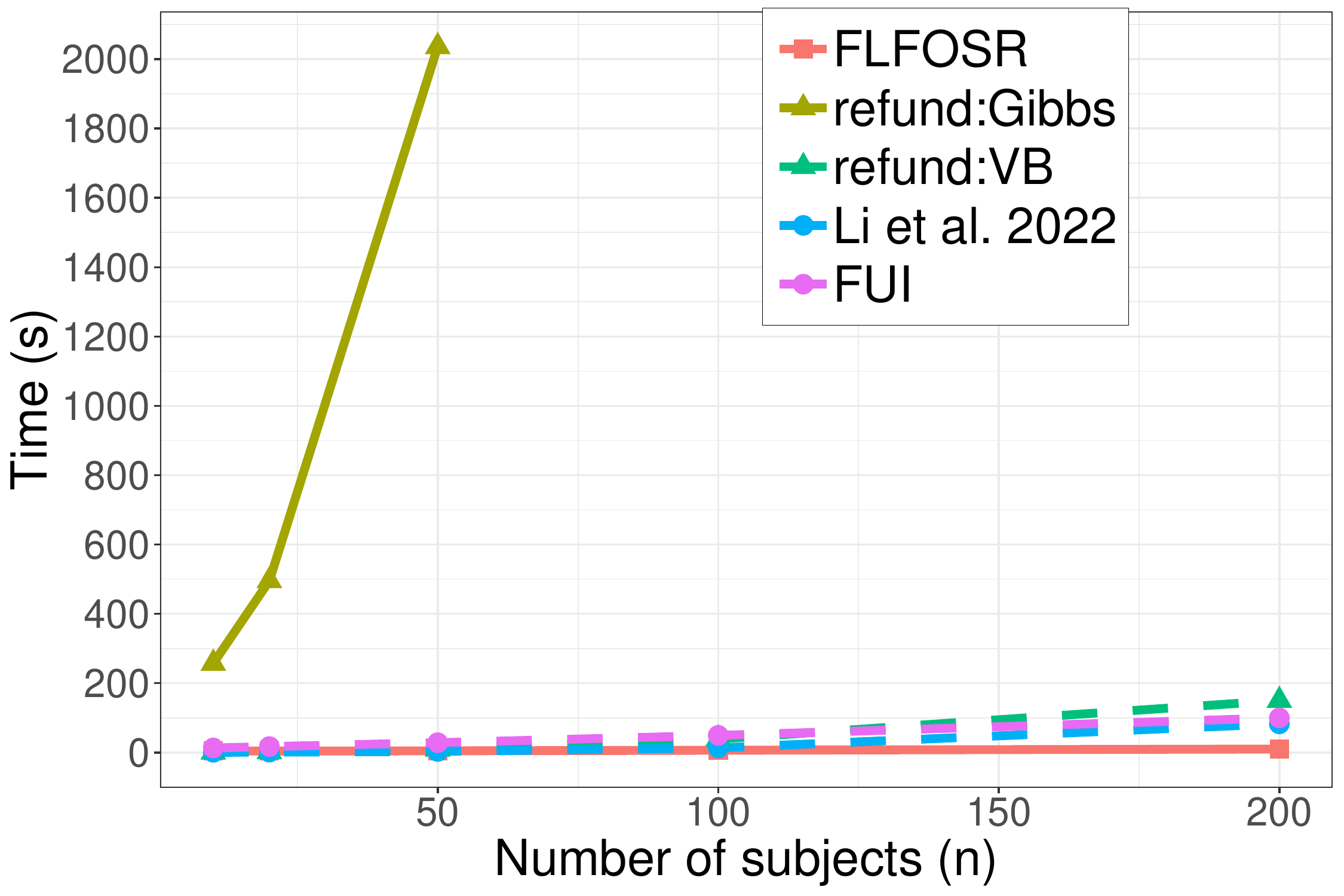}
\caption{}
\label{fig:sub1}
\end{subfigure}%
\begin{subfigure}{.5\linewidth}
\centering
\includegraphics[width=1\textwidth]{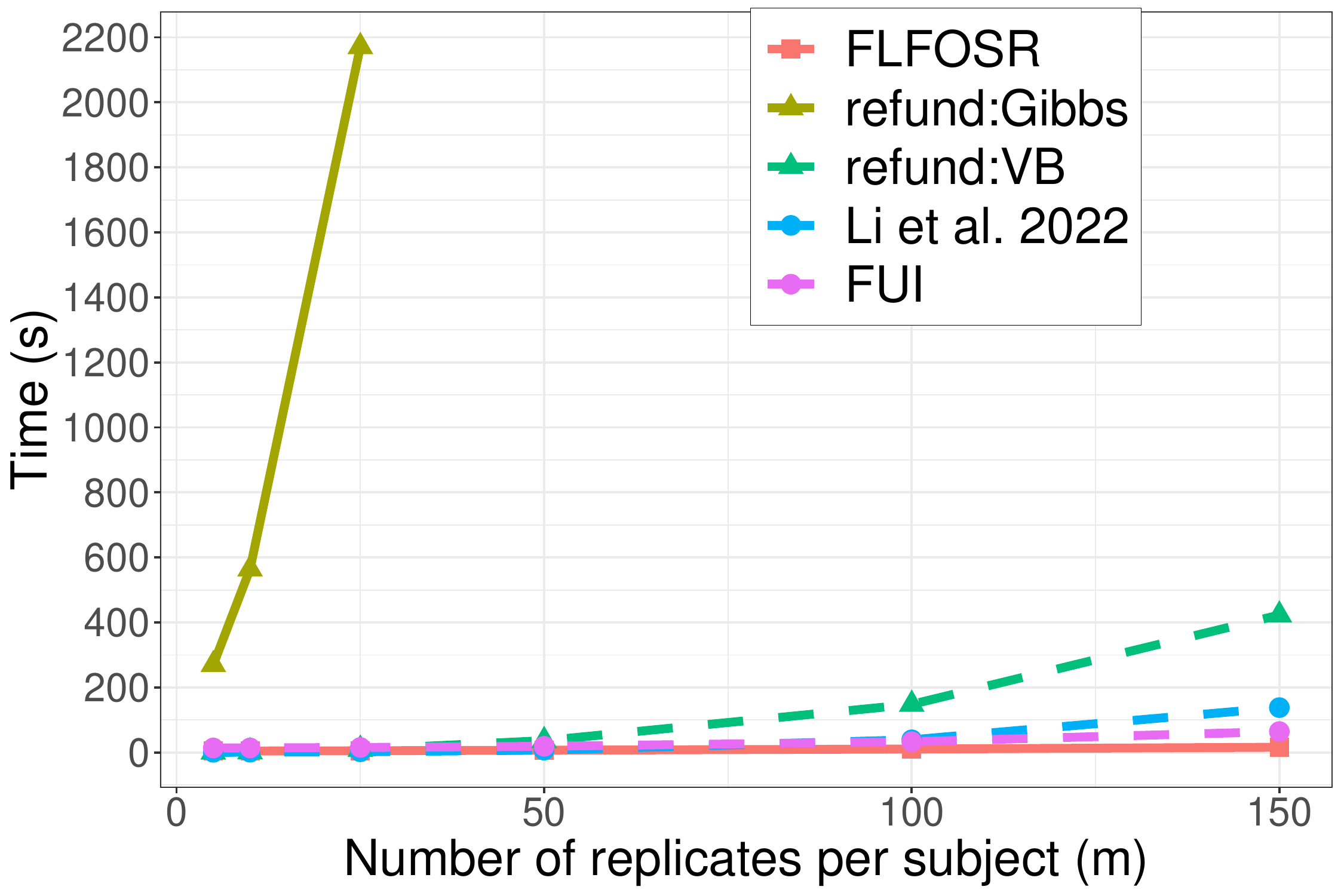}
\caption{}
\label{fig:sub2}
\end{subfigure}\\[1ex]
\begin{subfigure}{\linewidth}
\centering
\includegraphics[width=0.5\textwidth]{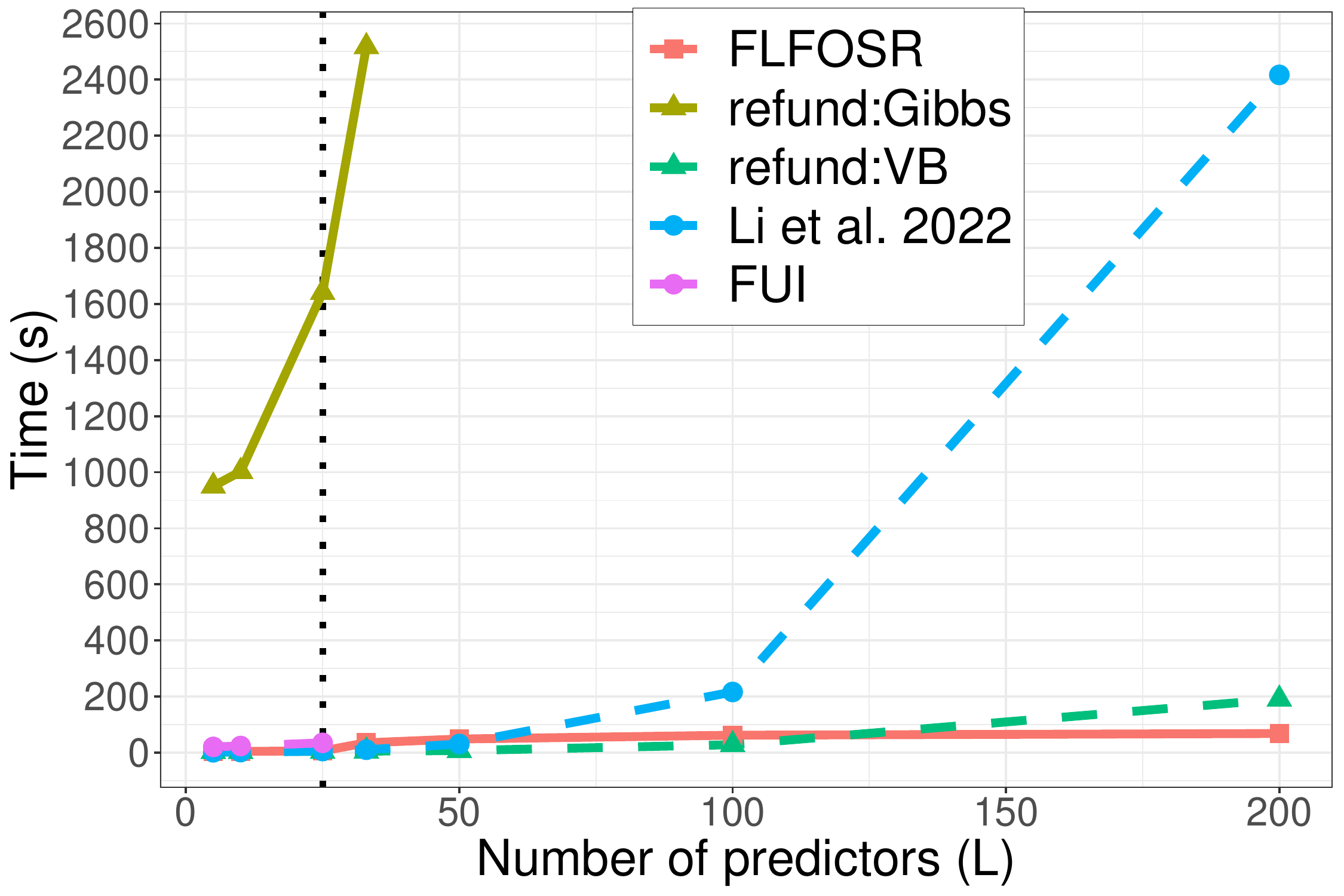}
\caption{}
\label{fig:sub3}
\end{subfigure}
\caption{Run time comparison of algorithms for longitudinal function-on-scalars regression, scaling by (a) the number of subjects $n$, (b) number of repeated observations per subject $m_i$, and (c) number of scalar predictors $L$, reported in seconds. Empirical computing time for MCMC methods (solid lines) refers to average time to 1000 effective samples, $\Bar{s}_{1000}$, and other methods (dashed lines) report raw run time. Computing times were averaged across 30 replicate simulations. The proposed FLFOSR outperforms all competitors as each dimension grows. \texttt{refund:Gibbs} was omitted for some designs due to lack of competitiveness and FUI requires $L < n$ (dotted black line in (c)).}
\label{fig:runtimes}
\end{figure}

First, we compute the \emph{average relative efficiency} across all covariates and observation points, 
\[
\Bar N_{\text{eff}}/N = (LT)^{-1} \sum_{\ell=1}^L \sum_{t=1}^TN_{\text{eff}}\{\tilde{\alpha}_\ell\left(\tau_{t}\right)\}/N,
\] 
which exclusively measures MCMC efficiency (not computing time). We compare this metric between the competing MCMC samplers, FLFOSR and \texttt{refund:Gibbs}, using the simulation design from 
Section~\ref{sec:simdesign}. Figure~\ref{fig:mcmceff} presents results across various settings for the variance components $\{\sigma_{\alpha_{}}^{2*}, \sigma^{2*}_{\gamma}, \sigma^{2*}_{\omega},\sigma^{2*}_{\epsilon}\}$ and $n=20$, $m = 5$, and $L=5$ (a smaller dataset is necessary for  \texttt{refund:Gibbs}; see Figure~\ref{fig:runtimes}).  FLFOSR is dramatically more efficient than \texttt{refund:Gibbs}, with consistently excellent performance across all setting. By comparison, the MCMC efficiency of \texttt{refund:Gibbs} deteriorates significantly whenever $\sigma^{2*}_{\gamma}$ or  $\sigma^{2*}_{\omega}$ is large. This result is unsurprising: for Gibbs samplers that alternate between drawing the fixed and the random effects from their respective full conditional distributions (e.g., \texttt{refund:Gibbs}), it is well-known that the  model parameterization (centered vs. noncentered) is critical for MCMC efficiency 
\citep{yu_center_2011}. In particular, the performance depends on the relative magnitudes of the variance components, which in practice are unknown. By comparison, FLFOSR samples the fixed and random effects functions \emph{jointly}, and thus requires no consideration of the parametrization---and  achieves excellent MCMC efficiency across a variety of settings.



Next, we summarize aggregate computational performance by measuring the time needed to generate $1000$ \emph{effective} samples, averaged over all covariates and observation points,
\[
\Bar{s}_{1000} = (LT)^{-1}\sum_{\ell=1}^L \sum_{t=1}^T 
\Big[
s_{N_{\text{burn}}} +  s_{N} \times \frac{1000}{N_{\text{eff}}\{\tilde{\alpha}_\ell(\tau_{t})\}}
\Big]
\]
which rewards both MCMC efficiency and computational scalability. 
This quantity allows some comparison between Bayesian and frequentist algorithms: although an MCMC algorithm may be run arbitrarily long to secure greater accuracy,  1000 effective samples is a reasonable and conservative target for general sampling-based inference. 

Using the simulation design from Section~\ref{sec:simdesign}, we scale the size of simulated datasets across three separate dimensions: (i) the number of subjects $n \in \{10, 20, 50, 100, 200\}$, fixing $m = 5$ and $L = 5$; (ii) the number of replicates or repeated observations per subjects, $m_i = m \in \{5, 10, 25, 50, 100, 150\}$, fixing $n=10$ and $L = 5$; and (iii) the number of predictors $L \in \{5, 10, 25, 33, 50, 100, 200\}$, fixing $n=30$ and $m = 5$. Data were generated using $\sigma_{\alpha_{}}^{2*} = \sigma^{2*}_{\gamma}= \sigma^{2*}_{\omega} = 1$, and $\sigma^{2*}_{\epsilon} = 10$. The computing times for each algorithm are averaged across 30 datasets in each simulation setting. 

The computational performance of each method is summarized  in Figure~\ref{fig:runtimes} and detailed further in Table \ref{tab:runtimes}. Once again,  FLFOSR decisively outperforms the MCMC competitor (\texttt{refund:Gibbs}): FLFOSR delivers more efficient exploration of the posterior while reducing computational costs by several orders of magnitude. These results highlight the mutual importance of (i) the \emph{joint} sampling step for $\{\alpha_{k,\ell}, \gamma_{k,i}, \omega_{k,i,j}\}$ and (ii) the careful construction of fast constituent sampling steps (Algorithm~\ref{alg:mcmc}).  For instance, FLFOSR generates 1000 average effective samples in under 10 seconds for a dataset with 144,000 data points ($T=144$, $n=200$, $m=5$). By comparison, \texttt{refund:Gibbs} requires several minutes for only $n=10$ subjects, and is not feasible even for moderately-sized datasets. 

Yet most surprisingly, FLFOSR attains 1000 effective samples in \emph{less time} than it takes for either the VB approximation or state-of-the-art frequentist algorithms \citep{li_fixedeffects_2022,cui_fast_2022} to terminate. These computing gains accelerate as each dimension $n$, $m$, or $L$ increases. Further, FUI scales poorly in $L$ and requires $L < n$, while FLFOSR has no such restrictions. These results were obtained from a single machine without utilizing any parallel processing. Thus, further scalability is attainable for FLFOSR.

\begin{table}[h]
\centering
\begin{tabular}{cccccccccc}
\toprule
\multicolumn{3}{c}{} & \multicolumn{2}{c}{FLFOSR} & \multicolumn{2}{c}{\texttt{refund:Gibbs}} & \multicolumn{1}{c}{\texttt{refund:VB}} & \multicolumn{1}{c}{Li et al.} & \multicolumn{1}{c}{FUI}\\
\cmidrule(rl){4-5} \cmidrule(rl){6-7}
$n$   & $m$ & $L$   & $\Bar{s}_{1000}$ & $\Bar N_{\text{eff}}/N$ & $\Bar{s}_{1000}$  & $\Bar N_{\text{eff}}/N$ & $s$  & $s$  & $s$ \\ \midrule
       10 & 5 & 5 & 4.5 & 0.59 & 257.1 & 0.57 & 0.7 & 0.4 & 12.9 \\ 
        20 & 5 & 5 & 4 & 0.73 & 495.5 & 0.71 & 2 & 0.8 & 16.9 \\ 
        50 & 5 & 5 & 4.5 & 0.86 & 2036.4 & 0.79 & 9.6 & 2.9 & 28.1 \\ 
        100 & 5 & 5 & 6.3 & 0.88 & - & - & 38.8 & 13.1 & 49.1 \\ 
        200 & 5 & 5 & 9.9 & 0.9 & - & - & 149.9 & 82.6 & 99.6 \\ \hline
        10 & 5 & 5 & 4.7 & 0.59 & 270 & 0.57 & 0.8 & 0.5 & 13.8 \\ 
        10 & 10 & 5 & 4.5 & 0.65 & 564 & 0.65 & 2 & 0.7 & 13.9 \\ 
        10 & 25 & 5 & 5.2 & 0.73 & 2171.4 & 0.76 & 9.7 & 2 & 15.3 \\ 
        10 & 50 & 5 & 6.9 & 0.75 & - & - & 36.3 & 7.2 & 19 \\ 
        10 & 100 & 5 & 10.5 & 0.79 & - & - & 147.9 & 38.9 & 33.1 \\ 
        10 & 150 & 5 & 16 & 0.79 & - & - & 422.7 & 137.9 & 64.2 \\ \hline
        30 & 5 & 5 & 4 & 0.81 & 950.2 & 0.75 & 3.9 & 1.2 & 20.3 \\ 
        30 & 5 & 10 & 4.3 & 0.77 & 1002.5 & 0.74 & 4 & 1.9 & 23.5 \\ 
        30 & 5 & 25 & 6.1 & 0.63 & 1641.1 & 0.62 & 4.6 & 5.5 & 34.8 \\ 
        30 & 5 & 33 & 35.1 & 0.55 & 2517.7 & 0.53 & 5.2 & 10.3 & - \\ 
        30 & 5 & 50 & 48.5 & 0.42 & - & - & 7.2 & 30.7 & - \\ 
        30 & 5 & 100 & 62 & 0.37 & - & - & 28.2 & 216.2 & - \\ 
        30 & 5 & 200 & 68.2 & 0.46 & - & - & 190.6 & 2417 & - \\ 
\bottomrule
\end{tabular}
\caption{Evaluating algorithm performance via time to 1000 effective samples ($\Bar{s}_{1000}$), average relative efficiency $(\Bar N_{\text{eff}}/N$), and raw computing time ($s$) (in seconds). The proposed MCMC algorithm (FLFOSR) is orders of magnitude faster than the MCMC competitor (\texttt{refund:Gibbs}) and even surpasses state-of-the-art VB and frequentist competitors \citep{li_fixedeffects_2022,cui_fast_2022} as each dimension grows. \texttt{refund:Gibbs} was omitted for some designs due to lack of competitiveness and FUI requires $L < n$.} 
  \label{tab:runtimes}
\end{table}

These results are extraordinarily favorable: Bayesian inference for \eqref{fmm} is especially appealing due to the availability of full posterior uncertainty quantification for all fixed and random effects functions and predictions, yet is often eschewed in favor of frequentist methods due to computational limitations. FLFOSR obviates this tradeoff, and  converts the computational performance from a disadvantage into a clear advantage for Bayesian FMMs.  




\subsection{Evaluating model accuracy and uncertainty quantification}\label{sec:simacc}

We assess the accuracy of each method by evaluating point and interval estimation for the fixed effect functions $\{\tilde\alpha_{\ell}(\cdot)\}$.  We measure point estimation accuracy using root mean squared error for the fixed effect functions,
\[
\text { RMSE }=\sqrt{\frac{1}{LT} \sum_{\ell=1}^L \sum_{t=1}^T\{\hat{\alpha}_\ell\left(\tau_{t}\right)-\tilde{\alpha}_\ell^*\left(\tau_{t}\right)\}^2}
\]
where $\hat{\alpha}_\ell\left(\tau_{t}\right)$ and $\tilde{\alpha}_\ell^*\left(\tau_{t}\right)$ are the estimated and true fixed effects functions, respectively, for the $\ell$th predictor at time $\tau_t$. For Bayesian methods, the estimator $\hat{\alpha}_\ell(\tau_{t})$ is the posterior mean. We evaluate uncertainty quantification using mean credible/confidence interval widths, 
\[
\text{MCIW} =\frac{1}{LT} \sum_{\ell=1}^L \sum_{\ell=1}^T\Big\{\tilde{\alpha}_{\ell}^{(.975)}\left(\tau_{t}\right)-\tilde{\alpha}_{\ell}^{(.025)}\left(\tau_{t}\right)\Big\}
\]
paired with empirical coverage probability, 
\[
\text{ECP} = \frac{1}{LT} \sum_{\ell=1}^L \allowbreak \sum_{t=1}^T \mathbb{I}\Big\{\tilde{\alpha}_{\ell}^{(.025)}\left(\tau_{\ell}\right) \leq \tilde{\alpha}_{\ell}^*\left(\tau_{t}\right) \leq \tilde{\alpha}_{\ell}^{(.975)}\left(\tau_{t}\right)\Big\},
\]
where $(\tilde{\alpha}_{\ell}^{(.025)}\left(\tau_{\ell}\right), \tilde{\alpha}_{\ell}^{(.975)}\left(\tau_{\ell}\right))$ are 95\% credible/confidence intervals. Ideal performance is achieved by nominal coverage $\mbox{ECP} \ge 0.95$ (calibration) and small MCIW (sharpness).

\begin{figure}[h]
\centering
\includegraphics[width=0.99\textwidth]{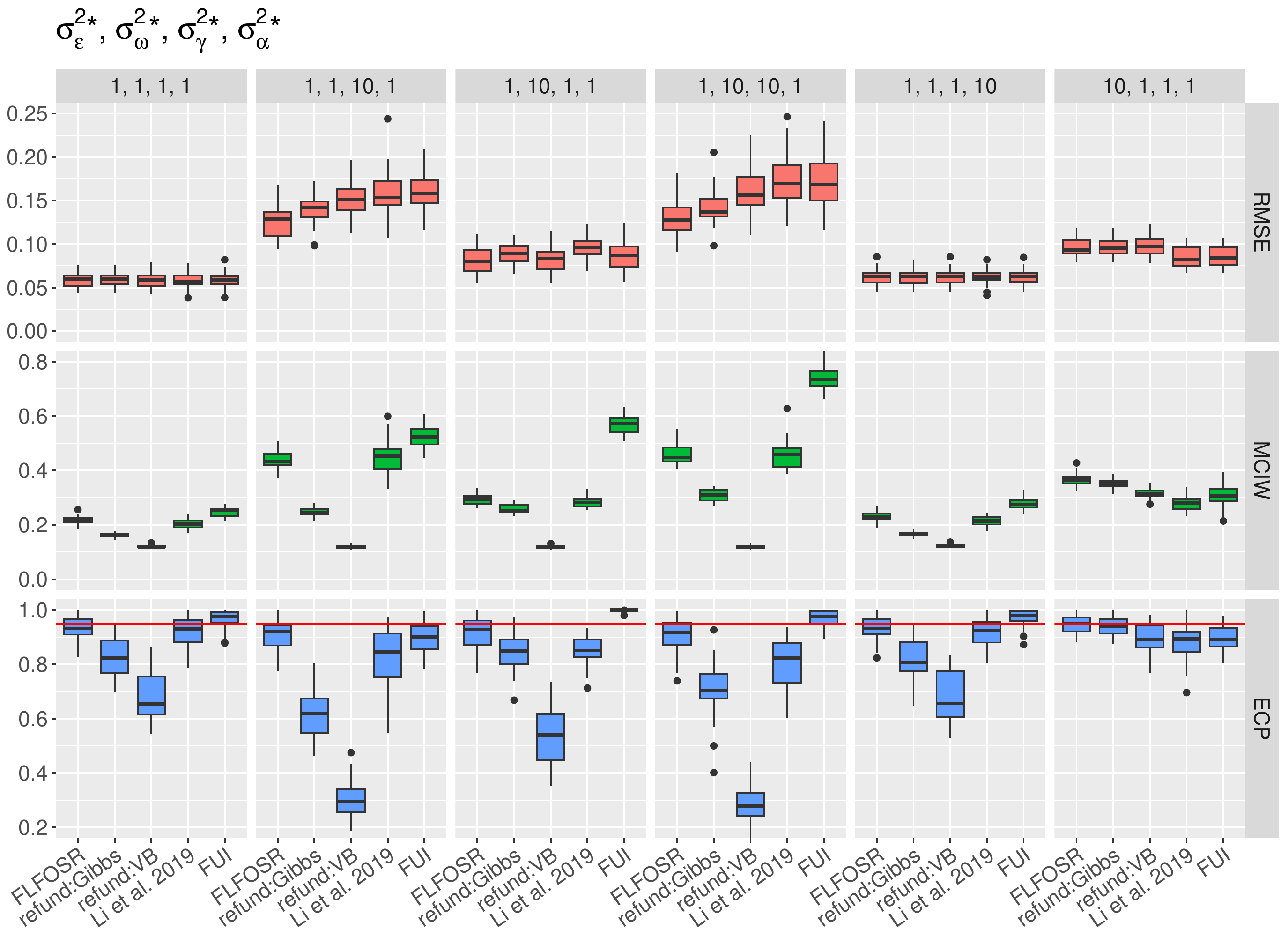}
\caption{Root mean squared errors (top row), mean credible/confidence interval widths (middle row), and empirical coverage probabilities (bottom row; red line denotes nominal coverage) for the fixed effects functions $\tilde \alpha_\ell(\cdot)$ across various designs (columns) and 30 simulated datasets (boxplots). The proposed FLFOSR consistently delivers accurate point estimation and precise and well-calibrated uncertainty quantification.}
  \label{fig:acc}
\end{figure}

We study the impact of differing sources of variability by varying the variance components $\{\sigma_{\alpha_{}}^{2*}, \sigma^{2*}_{\gamma}, \sigma^{2*}_{\omega},\sigma^{2*}_{\epsilon}\}$ between 1 and 10 and set $n=20$, $m=5$, and $L=5$. The results are summarized in Figure~\ref{fig:acc}. Broadly, the proposed FLFOSR is highly competitive across all scenarios for both point and interval estimation. Point estimation accuracy is comparable across all methods when between-subject variability is low ($\sigma_\gamma^{2*} = 1$), but FLFOSR offers substantial gains in point estimation accuracy as between-subject variability increases ($\sigma_\gamma^{2*} = 10$). Notably, FLFOSR achieves close to nominal coverage across all simulation designs, while \texttt{refund:Gibbs}, \texttt{refund:VB}, and \citet{li_fixedeffects_2022} suffer from significant undercoverage. Clearly, the VB approximation is reasonably accurate for point estimation but unreliable for uncertainty quantification. FUI is the only competitor that achieves close to nominal coverage, yet the FUI intervals are much wider than the FLFOSR intervals and thus sacrifices some power to detect important effects. Further, FUI is the least accurate point estimator when $\sigma_\gamma^{2*} = 10$. When measurement error dominates ($\sigma_\epsilon^{2*}=10)$, the Bayesian point estimates---including for FLFOSR---are less accurate. In such high-noise settings, this performance might be improved by replacing the priors \eqref{priors} with more aggressive shrinkage priors \citep{gao_bayesian_2022}.

In aggregate, the computational  (Figures~\ref{fig:mcmceff}--\ref{fig:runtimes}) and inferential  (Figure~\ref{fig:acc}) evaluations decisively favor FLFOSR. The proposed MCMC sampling algorithm (Algorithm~\ref{alg:mcmc}) is more scalable than existing frequentist methods and  more efficient than existing MCMC samplers, and delivers point and interval estimates that outperform state-of-the-art Bayesian and frequentist competitors. These significant gains amplify the additional benefits of Bayesian inference for FMMs, including  uncertainty quantification for all fixed and random effects functions and predictions---which was previously not accessible even for moderately-sized datasets. By comparison, frequentist approaches typically provide inference only for the fixed effects functions, and not for the random effects functions or predictions \citep{li_fixedeffects_2022,cui_fast_2022}.


\section{Application}\label{sec:app}
We apply FLFOSR  to analyze physical activity (PA) data measured from wearable accelerometry devices. These devices provide high resolution PA measurements for individual subjects across multiple days. PA plays a major role in overall human health, with lower PA levels linked to higher all-cause mortality  \citep{schmid_associations_2015, smirnova_predictive_2020}. However, statistical analysis of PA data often includes substantial aggregation or averaging both within a day and across days for a given subject \citep{kowal_subset_2022,hilden_how_2023}. Such pre-processing needlessly reduces the sample size and can obscure important sources of variability. In response, functional data analysis has become increasingly useful to analyze and model PA data \citep{sera_using_2017, leroux_organizing_2019, kowal_bayesian_2020, kowal_fast_2022}. Naturally, such approaches must confront the significant computational challenges associated with high resolution data for many individuals. 

Using data from the 2005-2006 NHANES cohort, we model PA as a function of time-of-day $\tau$ on days $j=1,\ldots,m_i$ for subjects $i=1,\ldots,n$. Importantly, this representation allows for analysis of time-of-day PA patterns via a FMM \eqref{fmm}, while also accounting for within-day autocorrelations, within-subject dependencies, and measurement errors. We follow the pre-processing procedures outlined in \citet{leroux_organizing_2019} and the accompanying \texttt{nhanesdata} \texttt{R} package, which removes subjects with poor data quality or too few days of measured activity, and restrict our analysis to subjects aged 35-85. The functional data $Y_{i,j}(\tau)$ are defined by computing the square-root of 10-minute PA averages over the window $\mathcal{T} = [\mbox{4:00}, \mbox{23:59}]$, which focuses on the waking hours and helps satisfy the FMM assumptions \eqref{fmm}. 
These functional data are paired with important health and demographic variables (Table~\ref{tab:variables}).  The continuous variables are standardized to have mean 0 and standard deviation 1. The final dataset has $n = 1723$ subjects with $\mbox{median}(\{m_i\}_{i=1}^n) = 6$ days of observations per subject, for a total of $M = 10372$ days of measurements, $T = 144$ measurements per day, 
and $L = 20$ predictors per subject.

\begin{table}[h]
\centering
\begin{tabular}{cc}
\toprule
Variable & Values \\ \midrule
\multicolumn{2}{l}{ \textit{ Response variable: }} \\ 
Activity Level & $[0, 440.0]$ \\ \midrule
\multicolumn{2}{l}{ \textit{Sociodemographic variables:} } \\ 
Gender & \textbf{Male} (53\%), Female (47\%) \\
Age (years) & $[35, 84.8]$ \\
\multirow{2}{*}{Race} & \textbf{White} (54\%), Black (21\%), \\
 & Hisp (22\%), Other (4\%) \\
Education Level & \textbf{$\boldsymbol{<}$ HS} (23\%), = HS (53\%), $>$ HS (24\%) \\ \midrule
\multicolumn{2}{l}{ \textit{Alcohol and drug use variables: }} \\ 
Drinks Per Week & {$[0,105]$} \\
Smoke Cigs & Current (20\%), Former, (30\%), \textbf{Never} (50\%)  \\ \midrule
\multicolumn{2}{l}{ \textit{Health-related variables:} } \\ 
Body Mass Index ($\mathrm{kg} / \mathrm{m}^2$) & {$[15.9,57.4]$} \\
HDL Cholesterol ($\mathrm{mg} / \mathrm{dL}$) & {$[23,188]$} \\
Total Cholesterol ($\mathrm{mg} / \mathrm{dL}$) & {$[92,458]$} \\
Systolic Blood Pressure (mmHg) & {$[80,270]$} \\
Has Congestive Heart Disease & Yes $(4 \%)$, \textbf{No} $(96 \%)$ \\
Has Congestive Heart Failure & Yes $(2 \%)$, \textbf{No} $(98 \%)$ \\
Has Cancer  & Yes $(9 \%)$, \textbf{No} $(91 \%)$ \\
Has Diabetes  & Yes $(10 \%)$, \textbf{No} $(90 \%)$ \\
Has Stroke  & Yes $(2 \%)$, \textbf{No} $(98 \%)$ \\ \midrule
\multicolumn{2}{l}{ \textit{Other variables:} } \\ 
Is Weekend & Yes $(28 \%)$, \textbf{No} $(72 \%)$ \\
\bottomrule
\end{tabular}
\caption{List of variables used in regression analysis of NHANES PA dataset and corresponding values. Baseline categories used for categorical variables are bolded.} 
  \label{tab:variables}
\end{table}

First, we compare the proposed FLFOSR approach with competing Bayesian methods  (\texttt{refund:Gibbs} and \texttt{refund:VB}). We adopt the same hyperparameter and MCMC settings as in Section~\ref{sec:sims} and conduct a sensitive analysis in the supplementary material. However, we are unable to run \texttt{refund:Gibbs} or \texttt{refund:VB} due to memory limitations. Instead, we fit these competing methods on a random subsample of $n=278$ subjects, the largest subsample our memory could accommodate, solely for the purpose of computational comparisons. 
Table \ref{tab:runtime_nhanes} shows the computational costs and MCMC efficiency for FLFOSR---fit to the whole dataset---compared to  \texttt{refund:Gibbs} and \texttt{refund:VB}.  FLFOSR required only about a minute to run, or about five minutes to generate 1000 effective samples from the posterior of the fixed effects functions. By comparison, \texttt{refund:Gibbs} required more than 21 hours to run, and due to MCMC inefficiencies, would require almost \emph{two weeks} to generate 1000 effective samples---even on this much smaller dataset. Similarly, the VB approximation is dramatically slower than the FLFOSR MCMC sampling algorithm, again on the smaller dataset. Thus, the \citet{goldsmith_assessing_2016} strategy of  (i) using VB to obtain initial estimates and (ii) using  Gibbs sampling for the final results is neither necessary nor feasible: we instead should proceed exclusively with FLFOSR, which provides fully Bayesian inference on the complete dataset in a fraction of the time.

\begin{table}[h]
\centering
\begin{tabular}{ccccc}
\toprule
Method & Dataset & Time ($N=1000$) & $\Bar{s}_{1000}$ & $\Bar N_{\text{eff}}/N$ \\ 
\hline FLFOSR & Full ($n=1723$) & 1.3 minutes & 4.8 minutes & .27 \\ 
\texttt{refund:Gibbs}  & Reduced ($n=278$) & 21.3 hours & 333.5 hours & .06 \\
\texttt{refund:VB} & Reduced ($n=278$) & 7.5 minutes & - & - \\
\bottomrule
\end{tabular}
\caption{Computational comparisons among Bayesian algorithms for the NHANES dataset. Performance is evaluated using time to $N=1000$ MCMC samples, time to 1000 effective samples ($\Bar{s}_{1000}$), and average relative efficiency $(\Bar N_{\text{eff}}/N$). The proposed FLFOSR requires only a fraction of the computing time for a much larger dataset.} 
  \label{tab:runtime_nhanes}
\end{table}

We summarize the FMM inference for select fixed effects functions $\tilde \alpha_\ell(\cdot)$ in  Figure~\ref{fig:act_coef} (see supplementary material for the remaining fixed effects functions).  The results show interesting, but not necessarily unsurprising relationships between certain health factors and time-of-day PA. Age is a strong negative predictor of activity, especially in the afternoon hours. Current cigarette smokers are less active, predominantly in the morning, but this effect is not as strong later in the day. Individuals with higher educational attainment are less active throughout the daytime, but more active after in the early evening. Among health variables, certain comorbidities like diabetes are associated with less activity. HDL cholesterol, commonly referred to as ``good cholesterol", is associated with higher activity throughout the day.

\begin{figure}[h]
\centering
\includegraphics[width=0.99\textwidth]{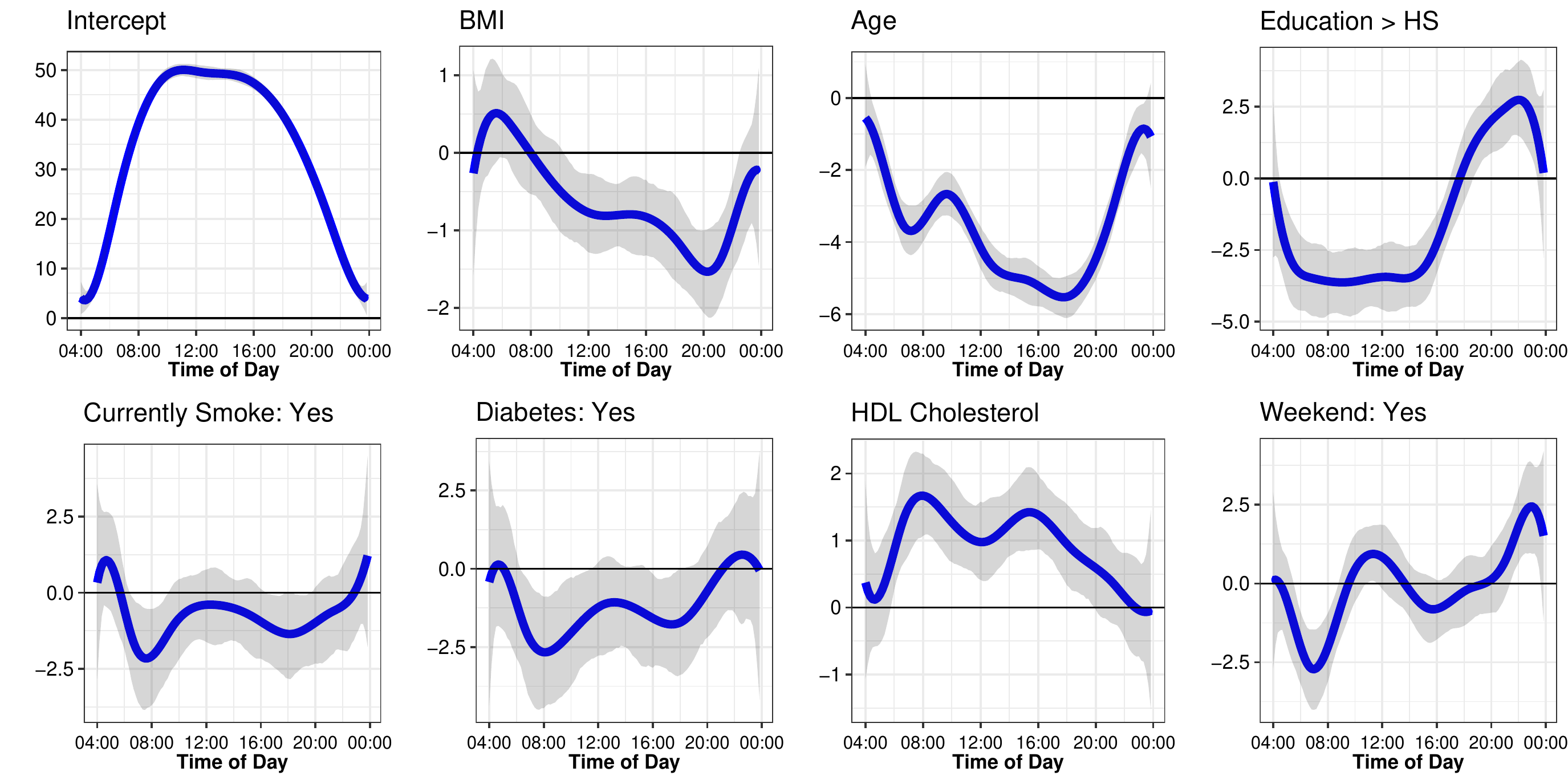}
\caption{Posterior inference for the fixed effects functions $\tilde \alpha_\ell(\cdot)$ for select covariates. The solid blue lines denote the posterior mean and the gray shading denotes the 95\% pointwise credible intervals. }
  \label{fig:act_coef}
\end{figure}

\begin{figure}
\begin{subfigure}{.5\linewidth}
\centering
\includegraphics[width=1\textwidth]{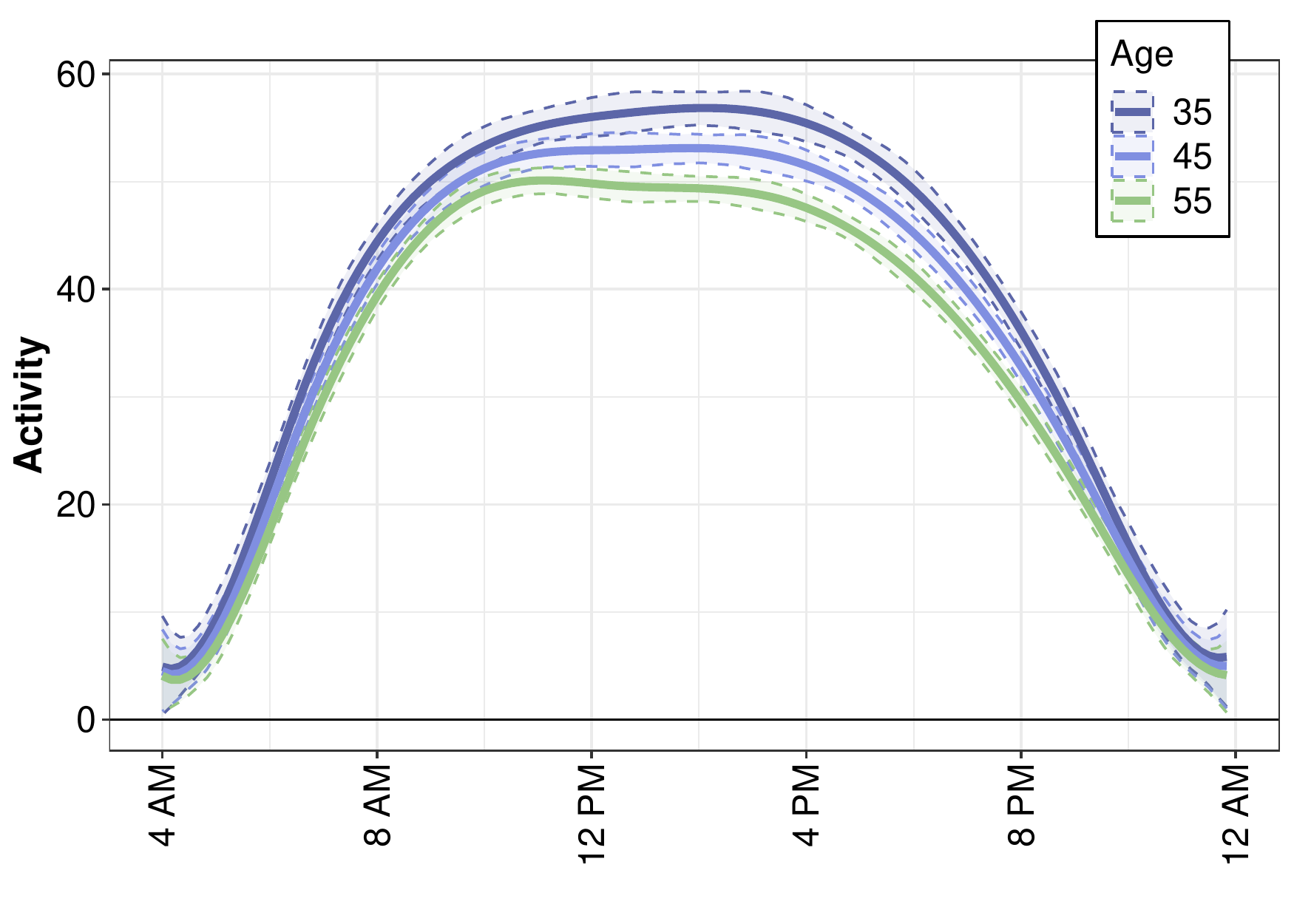}
\caption{}
\label{fig:actsub1}
\end{subfigure}%
\begin{subfigure}{.5\linewidth}
\centering
\includegraphics[width=1\textwidth]{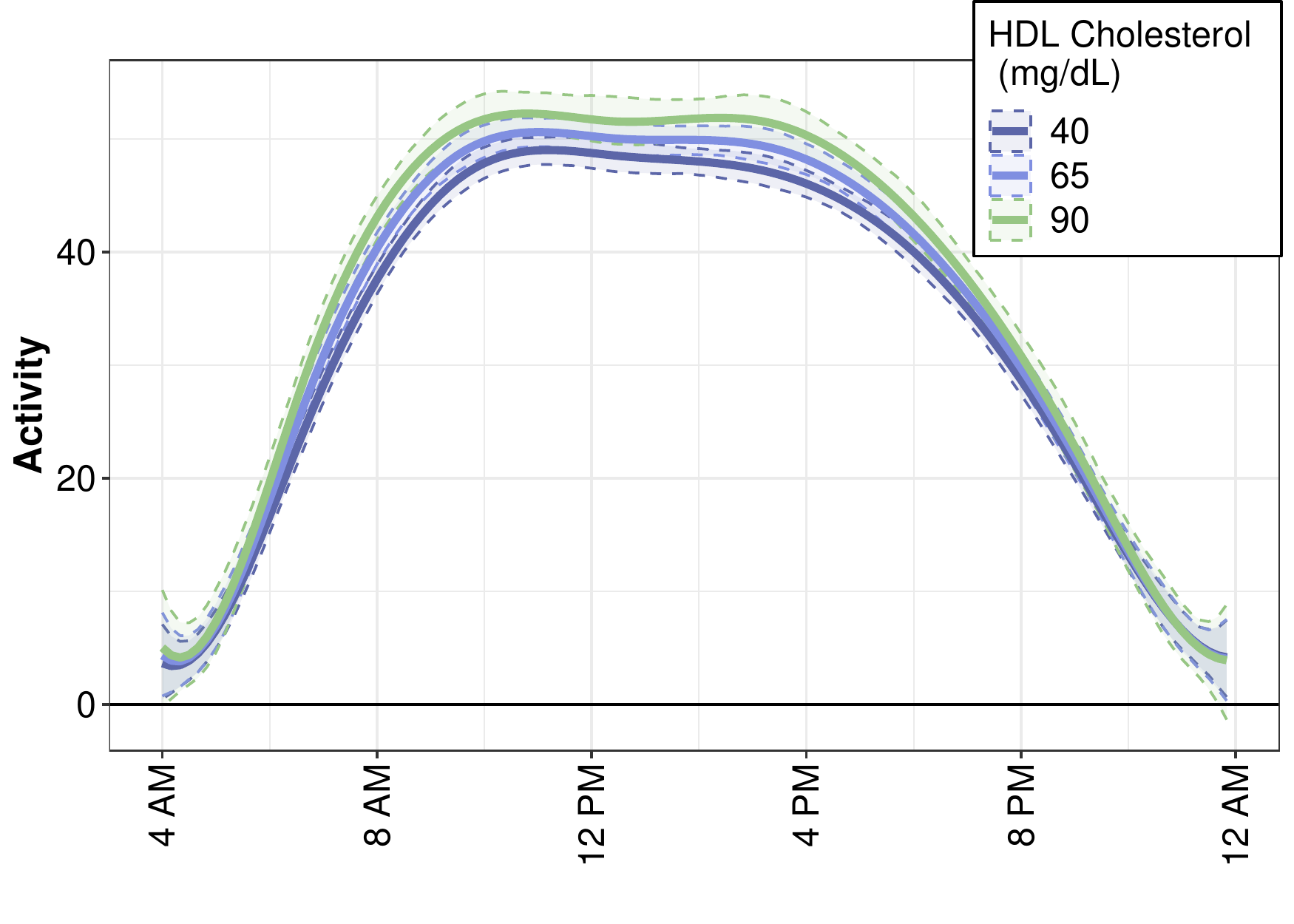}
\caption{}
\label{fig:actsub2}
\end{subfigure}\\[1ex]
\caption{Mean activity level for varying levels of (a) age and (b) HDL cholesterol with 95\% credible intervals for a subject with baseline categorical covariate levels and average values for continuous covariates.}
\label{fig:meanact}
\end{figure}

As an alternative visualization of these effects, Figure \ref{fig:meanact} shows the predicted population-level mean activity for varying levels of (a) age and (b) HDL cholesterol. The other predictors were set to the baseline and the mean values for categorical and continuous variables, respectively. As age increases from 35 to 55, there is a substantial overall decrease in predicted PA, which is particularly pronounced in the afternoon. HDL cholesterol levels in the data ranged from 23 mg/dL to 188 mg/dL, but levels below 40 mg/dL are broadly classified as ``at risk" of heart disease and are generally recommended to be above 60 mg/dL \citep{james_c_executive_2001}. The right hand plot shows the predicted activity curves when raising HDL levels from the lower cutoff point up into the recommended amount, where slightly heightened PA throughout waking hours is predicted as HDL levels increase. 

\section{Discussion}
\label{sec:disc}
We introduced a new MCMC algorithm for Bayesian regression with longitudinal functional data. The algorithm applies for a broad and widely-useful class of functional mixed models that include (i) basis expansions and (ii) fixed effects and subject- and replicate-specific random effects. Most notably, our MCMC sampler delivers unrivaled MCMC efficiency and unmatched computational scalability, and is empirically faster than state-of-the-art frequentist competitors---while also providing posterior uncertainty quantification for all model parameters and predictions. 
The scalability of our algorithm persists across all dimensions---the number subjects, replicates per subjects, and covariates---and thus establishes a new state-of-the-art for (Bayesian or frequentist) computing with longitudinal functional data. The proposed model showcases excellent point estimation accuracy and uncertainty quantification across a broad array of simulated data scenarios.  

In practice, our MCMC algorithm enables complex Bayesian model-fitting with large functional datasets without the need for intensive computing resources. Indeed, we were able to perform fully Bayesian inference on a large physical activity (PA) dataset in about a minute using a personal desktop without any parallel processing. As the usage of wearable devices continues to grow, so does the need for highly scalable methods to study these high-resolution and high-dimensional data. For example, the UK Biobank study contains PA measurements of over 100,000 subjects across 7 days, measured in 5 second intervals \citep{doherty_large_2017}, implying over 12 billion rows of PA measurements. 


To meet these computational burdens, future work will consider EM algorithms and VB approximations based on the same core ideas as the proposed MCMC algorithm (Algorithm~\ref{alg:mcmc}). We expect that such strategies would further increase computational scalability, but at the cost of reliable (posterior) uncertainty quantification. 


One computational aspect that we did not explore in this paper is the memory usage of each algorithm. For the NHANES data, memory issues precluded use of the competing methods in the \texttt{refund} package on the whole dataset, while the proposed FLFOSR approach encountered no such problems. Thus, efficient memory usage may be another advantage of FLFOSR, but further analysis is needed. 


\bigskip
\begin{center}
{\large\bf SUPPLEMENTARY MATERIAL}
\end{center}

\begin{description}

\item[Additional results:] A document containing a sensitivity analysis and additional results for the application. (PDF)

\item[R code:] Code for model implementation and producing all results and plots in the paper. (Zipped file)

\end{description}

\bibliography{mybib052723.bib}

\newpage

\supplementarysection
\renewcommand{\thesubsection}{\Alph{subsection}}

\subsection{Additional Simulation Results}
 \doublespacing

\begin{figure}[h]
\centering
\includegraphics[width=0.99\textwidth]{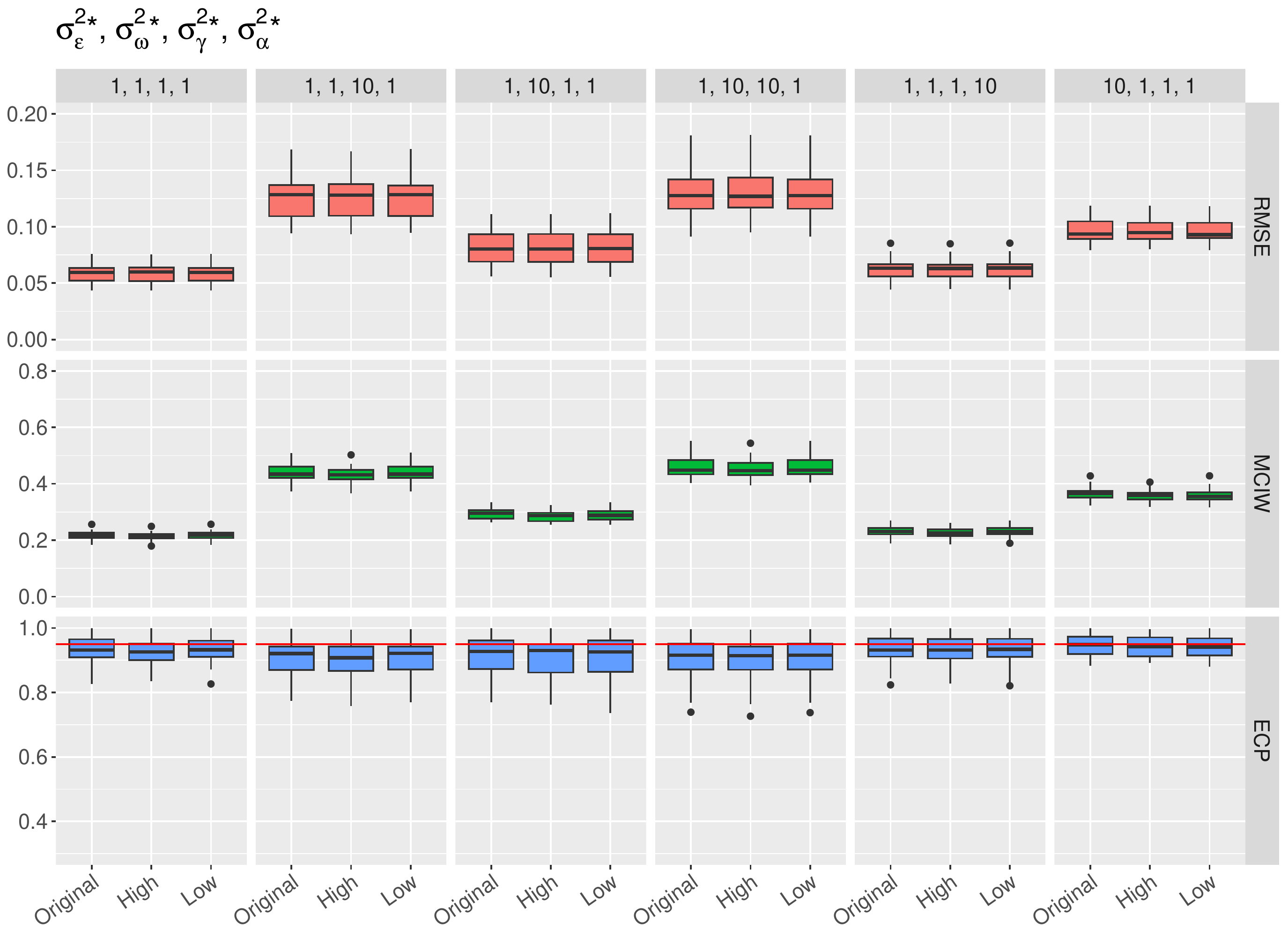}
\caption{Sensitivity analysis of FLFOSR results shown in Figure 4 for original, high and low values of variance hyperparameters. Root mean squared errors (top row), mean credible/confidence interval widths (middle row), and empirical coverage probabilities (bottom row; red line denotes nominal coverage) for the fixed effects functions $\tilde \alpha_\ell(\cdot)$ across various designs (columns) and 30 simulated datasets (boxplots).}
  \label{fig:acc_sens}
\end{figure}

We perform a sensitivity analysis to investigate how choice of hyperparameter values for the inverse-gamma priors on $\sigma_{\gamma_{}}^2$ and $\sigma_{\omega_{i}}^2$ affects the simulation results. We set hyperparameter values as $a_{\sigma^{}_{\gamma}} = 5$, $b_{\sigma^{}_{\gamma}} = 1$, $a_{\sigma^{}_{\omega}} = 5$, $b_{\sigma^{}_{\omega}} = 1$ as high values, and $a_{\sigma^{}_{\gamma}} = .005$, $b_{\sigma^{}_{\gamma}} = .001$, $a_{\sigma^{}_{\omega}} = .005$, $b_{\sigma^{}_{\omega}} = .001$ as low values. Figure \ref{fig:acc_sens} shows the original results of our model from the simulation study in Section 4.4 along with the same simulations but with high and low hyperparameter values. The results show generally low sensitivity to choice of variance hyperparameters. 

\subsection{Additional Application Results}

For the results from our data application in Section 5, Figure \ref{fig:act_s} shows the rest of the fixed effect functions for model covariates not shown in Figure 5. Certain health conditions such as cancer, stroke, and congestive heart failure seem to have close to zero effect on activity throughout the day, along with a large uncertainty band. Although, congestive heart disease may have a meaningful negative association in the morning and afternoon. Only a small percentage of the sample reported having any of these conditions (Table 2). Race and gender seem to be strong predictors of physical activity, whereas other effects were not a strong, such as drinks per week or being a former smoker (compared to non-smoker).

\begin{figure}
\begin{subfigure}[b]{1\textwidth}
\centering
\includegraphics[width=1\linewidth]{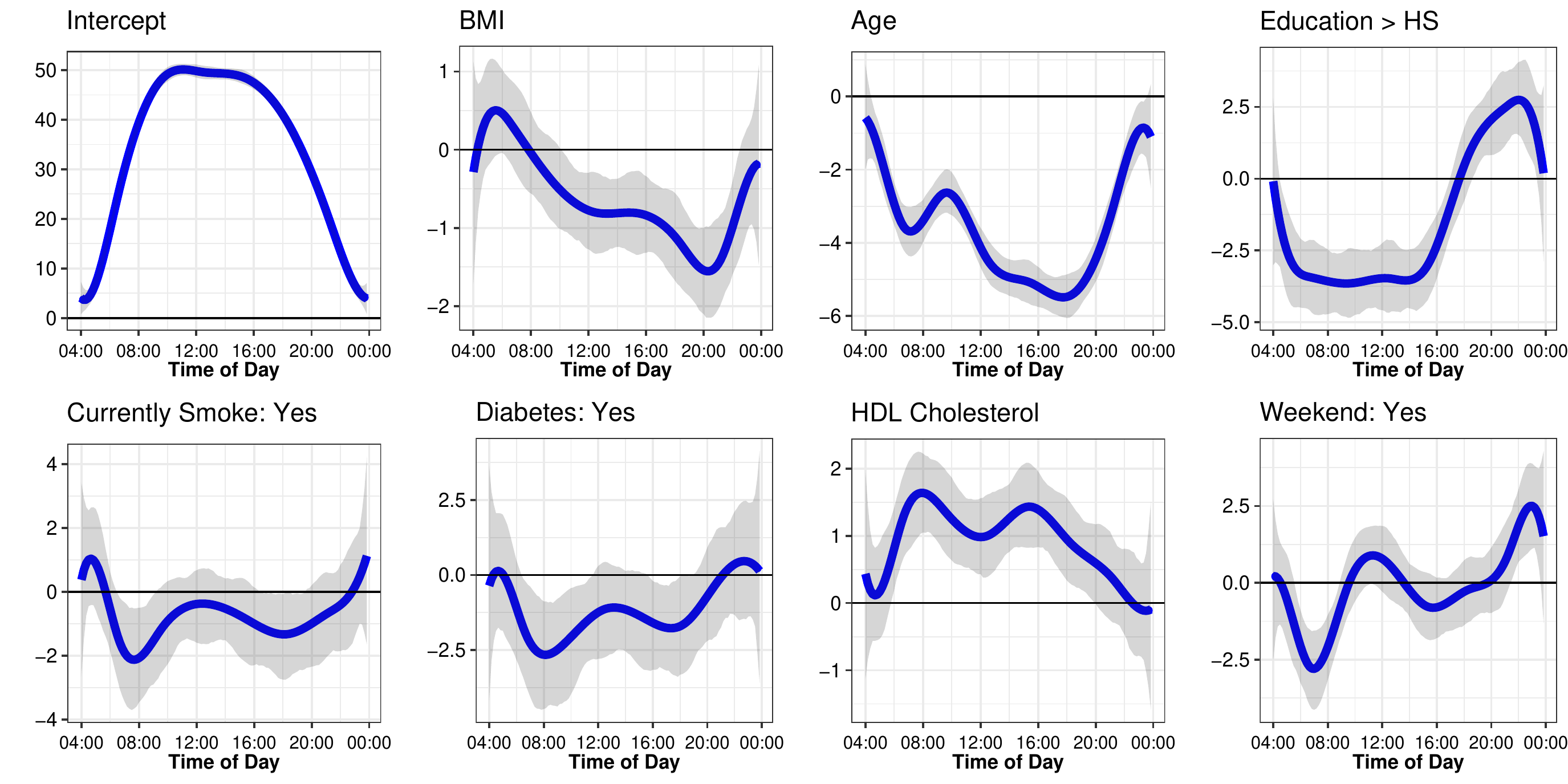}
\caption{}
\label{fig:actsens_sub1}
\end{subfigure}

\begin{subfigure}[b]{1\textwidth}
\centering
\includegraphics[width=1\linewidth]{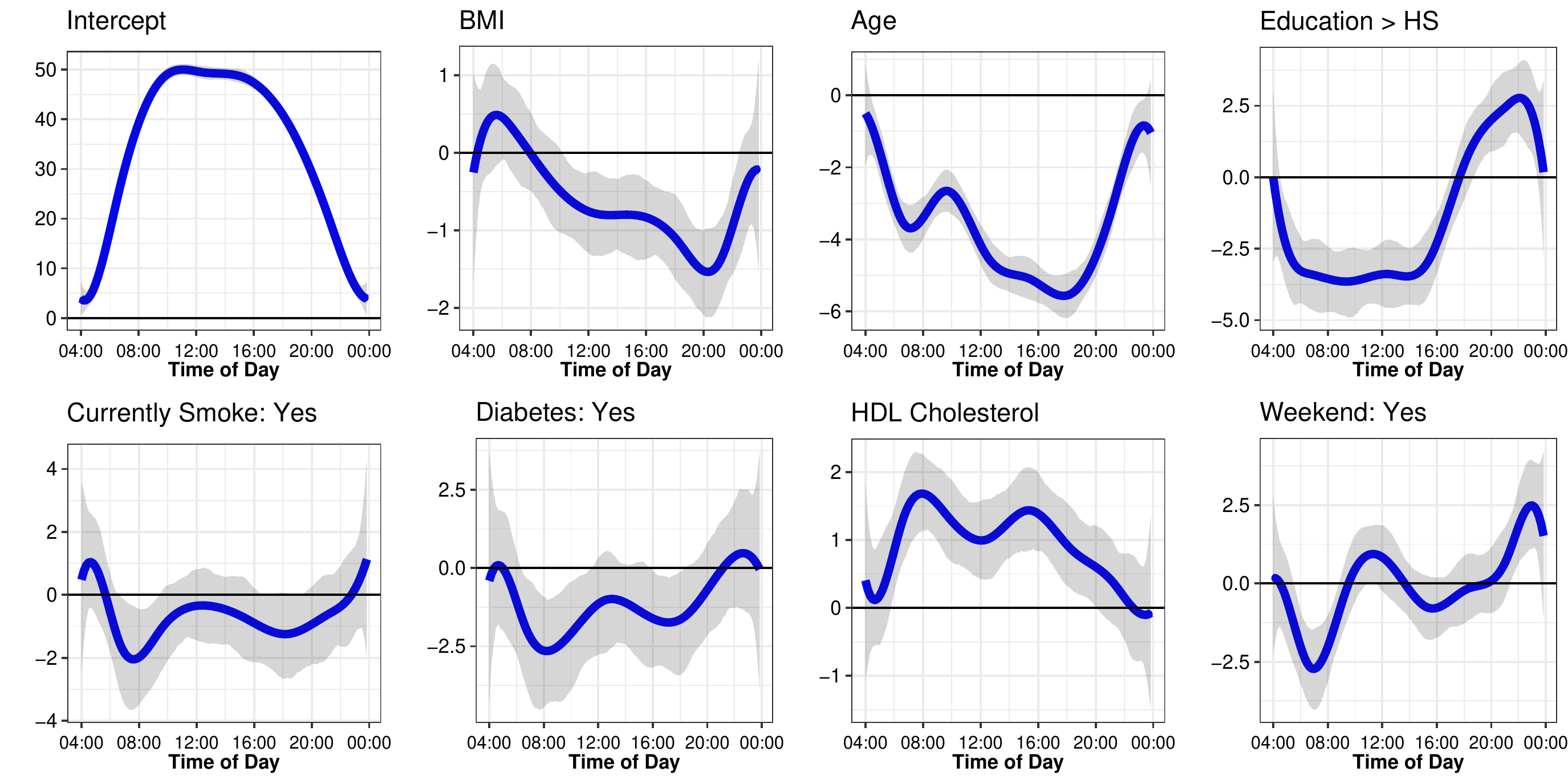}
\caption{}
\label{fig:actsens_sub2}
\end{subfigure}
\caption{Sensitivity analysis of results shown in Figure 5 for (a) high and (b) low values of variance hyperparameters. Results are almost identical to each other and the original results.}
\label{fig:act_sens}
\end{figure}

\begin{figure}[h]
\centering
\includegraphics[width=0.99\textwidth]{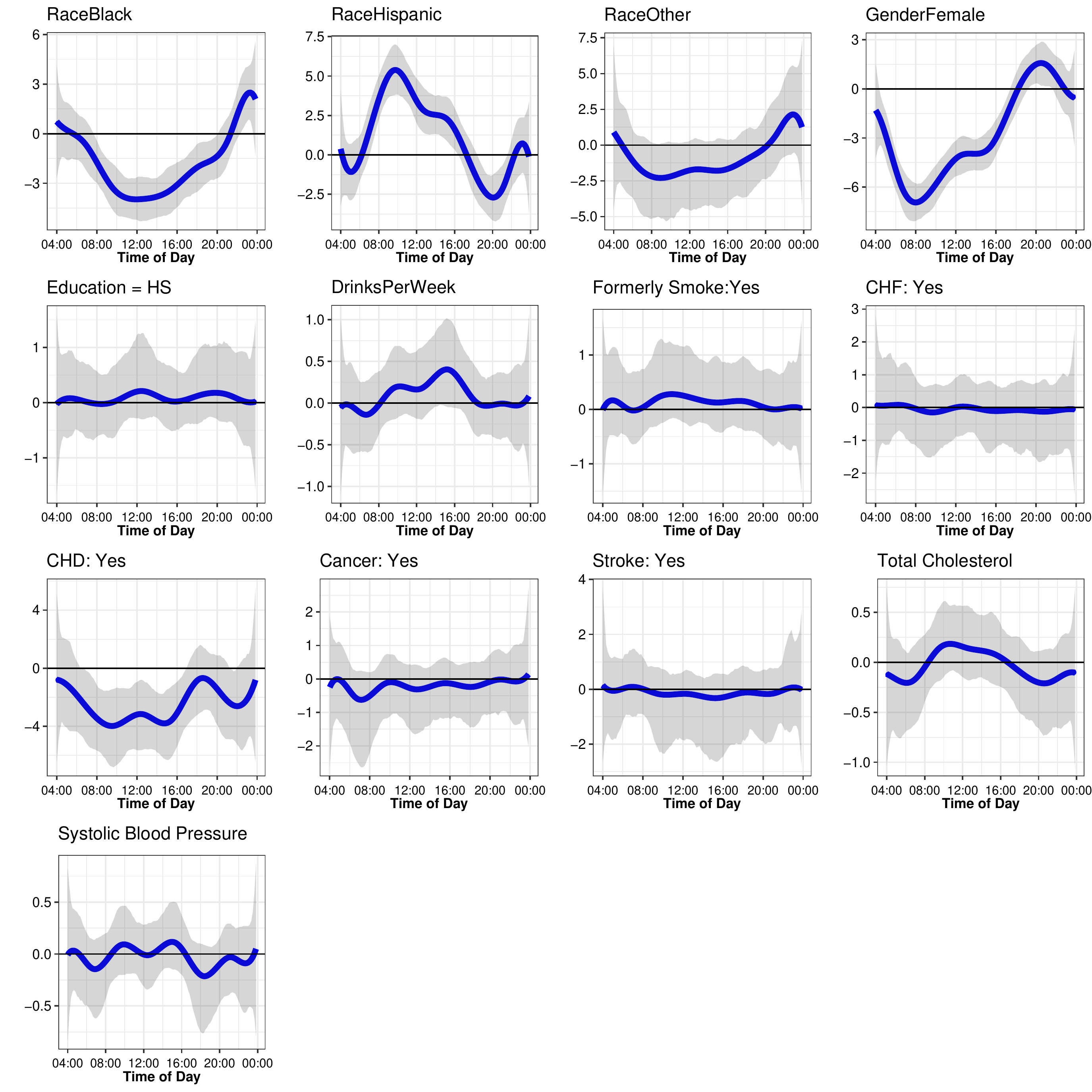}
\caption{Posterior inference of fixed effects functions for the remaining covariates not shown in Figure 5.}
  \label{fig:act_s}
\end{figure}

Figure \ref{fig:act_sens} shows a sensitivity analysis of the estimated fixed effect regression functions and 95\% credible intervals akin to Figure 5. Again, we set hyperparameter values as $a_{\sigma^{}_{\gamma}} = 5$, $b_{\sigma^{}_{\gamma}} = 1$, $a_{\sigma^{}_{\omega}} = 5$, $b_{\sigma^{}_{\omega}} = 1$ as high values, and $a_{\sigma^{}_{\gamma}} = .005$, $b_{\sigma^{}_{\gamma}} = .001$, $a_{\sigma^{}_{\omega}} = .005$, $b_{\sigma^{}_{\omega}} = .001$ as low values. The results are almost identical to each other and the original estimates, showing robustness to choice of hyperparameters. This is likely in part due to the large sample size of the NHANES dataset.

\end{document}